\documentclass[]{amsart}

\usepackage{epsfig}
\usepackage{graphics}
\usepackage{graphicx}
\usepackage{amsmath}
\usepackage{amsfonts}
\usepackage{amsthm}
\usepackage{amssymb}
\usepackage{url}
\usepackage{subfigure}
\usepackage{multicol}
\usepackage{captcont}

\newtheorem{finding}{Finding}

\begin{document}

\title[Slime mould imitation of Belgian motorways]{Slime mould imitation of \\ Belgian transport networks:\\ redundancy, bio-essential motorways, and dissolution}

\author[Adamatzky]{Andrew Adamatzky}
\address[Andrew Adamatzky]{Unconventional Computing Centre, University of the West of England, Bristol, United Kingdom}

\author[De Baets]{Bernard De Baets}
\address[Bernard De Baets]{Department of Mathematical Modelling, Statistics and Bioinformatics,
Ghent University, Gent 9000, Belgium}

\author[Van Dessel]{Wesley Van Dessel}
\address[Wesley Van Dessel]{Scientific Institute of Public Health,  Brussels B1050, Belgium}

\date{\today}


\begin{abstract}

\vspace{0.5cm}

Belgium is amongst few artificial countries, established on purpose, when Dutch and French speaking parts were joined in a single unit. This makes Belgium a particularly interesting testbed for studying bio-inspired techniques for simulation and analysis of vehicular transport networks.  We imitate growth and formation of a transport network between major urban areas in Belgium using the acellular slime mould \emph{Physarum polycephalum}. We represent the urban areas with the sources of nutrients.  The slime mould spans the sources of nutrients with a network of protoplasmic tubes. The protoplasmic tubes  represent the motorways. In an experimental laboratory analysis we compare the motorway network approximated by \emph{P. polycephalum} and the man-made motorway network of Belgium. We evaluate the efficiency of the slime mould network and the motorway network using proximity graphs.

\vspace{0.5cm}

\noindent
\emph{Keywords: transport networks, unconventional computing, slime mould} 
\end{abstract}

\maketitle

\section{Introduction}

Plasmodium is a vegetative stage of the acellular slime mould \emph{Physarum polycephalum}. This is a 
single cell with  many nuclei. The plasmodium feeds on microscopic particles~\cite{stephenson_2000}. 
During its foraging behaviour the plasmodium spans scattered sources of nutrients with a network of  protoplasmic tubes. The protoplasmic network is optimised to cover all sources of food and to provide a robust and speedy transportation of nutrients and metabolites in the plasmodium body. The plasmodium's foraging behaviour can be
interpreted as computation. Data are represented by spatial configurations of attractants and repellents, and  results of computation by structures of protoplasmic network formed by the plasmodium on the data sites~\cite{nakagaki_2000,nakagaki_2001a,PhysarumMachines}. The problems solved by plasmodium of \emph{P. polycephalum} include shortest path~\cite{nakagaki_2000, nakagaki_2001a}, 
implementation of storage modification machines~\cite{PhysarumMachines},
Voronoi diagram~\cite{shirakawa},  Delaunay triangulation~\cite{PhysarumMachines}, 
logical computing~\cite{tsuda_2004, adamatzky_gates}, and 
process algebra~\cite{schumann_adamatzky_2009}; see an overview in~\cite{PhysarumMachines}.
Previously~\cite{adamatzky_UC07} we have evaluated the road-modelling potential of \emph{P. polycephalum},
however no conclusive results were presented back in 2007.  A step towards biological approximation,
or evaluation, of man-made road networks was done in our previous papers on the approximation
of motorways/highways in the United Kingdom~\cite{adamatzky_jones_2009},
Mexico~\cite{adamatzky_Mexico} and Australia~\cite{adamatzky_Australia} by plasmodium of  \emph{P. polycephalum}. 
For these countries we found that, in principle, the network of protoplasmic tubes developed by plasmodium matches, at least partly, the network of man-made transport networks. However a country's shape and spatial configuration of urban areas, which are experimentally represented by sources of nutrients, might play a key role in determining the exact structure of plasmodium networks.  Also  we suspect that the degree of matching between the Physarum networks and the motorway networks is determined by original government designs of motorways in any particular country. This is why it is so important to collect data on the development of plasmodium networks in all major countries, and then undertake a comparative analysis.

Belgium is a good testbed for the evaluation of slime-mould approximation of motorways because 
\begin{itemize}
\item Belgium is an artificial country created relatively recently, in 1830.
\item It is amongst the most populated area in Europe.
\item There is a density misbalance between two major communities: Flanders is more densely populated than Wallonia.
\item the Belgian economy is centred around Brussels,  by far biggest city, with hundreds of thousands of workers commuting to Brussels every day.
\end{itemize}
In the early days the Belgian highways were constructed to provide a solution against the overcharged national and local roads, caused by the expanding number of cars. In the North of the country, construction was generally based on growing demands from the economic and touristic sectors. The first highway was the one between Brussels and Ostend. Another `early' highway was the one between Antwerp and Li\`{e}ge (E313) to open up the port of Antwerp's access to the `hinterland'. At the end of 1972 the most important cities were connected by highways. However, from the point of view of transport economics, only two of them were answering to an economic demand, a need for construction based on increasingly busy roads: the one between Brussels and Antwerp (E19), and the one between Brussels and Li\`{e}ge (E40). The others were intended as an investment trigger. The Autoroute de Wallonie (E42) was aimed at the economic reconversion of the old industrial axis in Wallonia (steel and coal industry). The E17 and E34 motorways provided an additional connection between the port of Antwerp and the French and German inner lands. The purpose of the E314 was to open up the province of Limburg (coal mining industry), and to provide a shortcut between Antwerp and the German Rhineland (Ruhrgebiet). Highway construction has been the result of political negotiations and the desire or need of the Northern and Southern partners to balance large investments in both parts of the country~\cite{WegenRoutes}.

The paper is structured as follows. In Sect.~\ref{methods} we give an  overview of the experimental setup employed.
Analysis of protoplasmic networks produced by slime mould \emph{P. polycephalum} in laboratory experiments 
is provided in Sect.~\ref{experimentalresults}. We compare slime mould generated and man-built motorways 
in Sect.~\ref{comparingPhysarumMotorways}. Section~\ref{proximity} considers protoplasmic networks and Belgian motorways in the context of planar proximity graphs. Relations between the experimental results and the administrative subdivision of Belgium are discussed in 
Sect.~\ref{adminstrativesubdivision}. In Sect.~\ref{contamination} we study outcomes of large-scale contamination and resulting 
reconfiguration of slime mould transport networks.

\section{Experimental}
\label{methods}

Plasmodium of \emph{ P. polycephalum} is cultivated in plastic container, on paper kitchen towels moistened with 
still water, and fed with oat flakes. For experiments we use $120 \times 120$~mm polystyrene square Petri dishes
and 2\% agar gel (Select agar, by Sigma Aldrich) as a substrate. Agar plates, about 2-3~mm in depth, are cut in 
the shape of Belgium.

\begin{figure}[!tbp]
\centering
\subfigure[]{\includegraphics[width=0.8\textwidth]{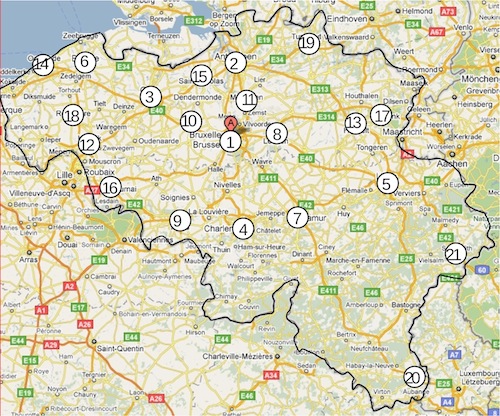}}
\subfigure[]{\includegraphics[width=0.8\textwidth]{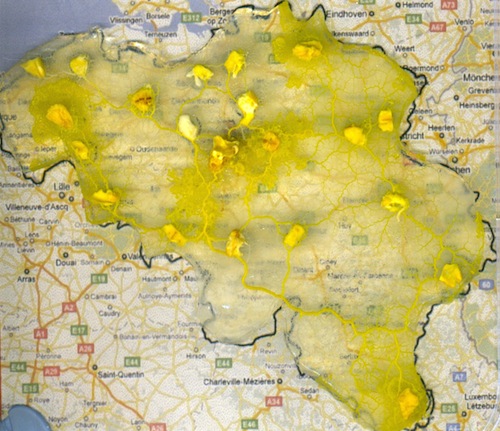}}
\caption{Experimental setup: (a)~outline map of Belgium~\cite{MapsGoogle} with major urban areas 
$\mathbf{U}$ shown by encircled numbers; (b)~urban areas, represented by oat flakes, are colonised by 
slime mould of \emph{P. polycephalum}. }
\label{urbanareas}
\end{figure}

We consider the twenty-one most\footnote{Arlon and Sankt-Vith are not amongst the most populated areas but we added them for completeness, to allows the slime mould propagating towards Luxembourg and Germany} populous urban areas in Belgium $\mathbf U$ (Fig.~\ref{urbanareas}a), shown below in descending order of population size:

\begin{multicols}{2}
\begin{enumerate}
\item Brussels area, including Dilbeek and Vilvoorde 
\item Antwerp area, including Beveren and Brasschaat 
\item  Gent 
\item Charleroi area, including La Louvi\`{e}re and Chatelet
\item Li\`{e}ge area, including Seraing, Verviers and Herstal
\item Brugge
\item Namur
\item Leuven
\item Mons 
\item Aalst
\item Mechelen
\item Kortrijk area, including Mouscron and Waregem 
\item Hasselt 
\item  Oostende 
\item Sint-Niklaas 
\item Tournai 
\item Genk area, including Maasmechelen 
\item Roeselare 
\item Turnhout 
\item Arlon 
\item Sankt-Vith 
\end{enumerate}
\end{multicols}

\begin{figure}[!tbp]
\centering
\subfigure[]{\includegraphics[width=0.8\textwidth]{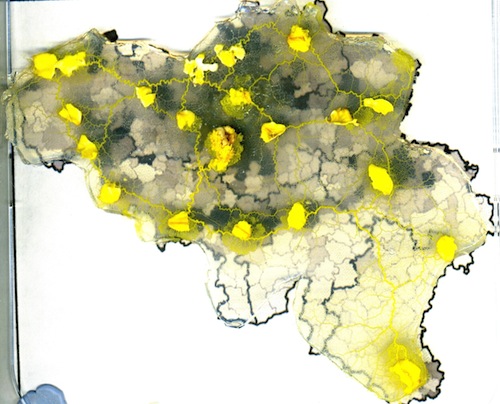}}
\subfigure[]{\includegraphics[width=0.8\textwidth]{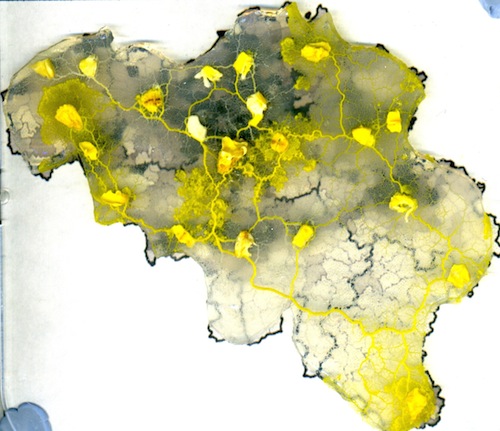}}
\caption{Oat flakes, representing urban areas $\mathbf{U}$, colonised by slime mould. The growing substrate 
is on top of the Belgian population density map~\cite{PopulationMap}.
}
\label{mappopulation}
\end{figure}

To represent areas of $\mathbf{U}$ we place oat flakes in the positions of agar plate corresponding to the areas. At the beginning of each experiment an oat flake colonised by plasmodium is placed in the Brussels 
area (Fig.~\ref{mappopulation}). Our choice of inoculation site  does not reflect the historical development of transport routes in Belgium (where inoculation should start in  Aalst, Brugge, Kortrijk, or Gent) however it conveys the overwhelming economic power of the capital.  We undertook 28 experiments. The Petri dishes with plasmodium are kept in darkness, at temperature 22-25$^\text{o}$C, except for observation and image recording. Periodically, the dishes are scanned with an Epson Perfection 4490 scanner.

\section{Slime mould transport networks: bio-essential motorways}
\label{experimentalresults}

\begin{figure}[!tbp]
\centering
\subfigure[24~h]{\includegraphics[width=0.45\textwidth]{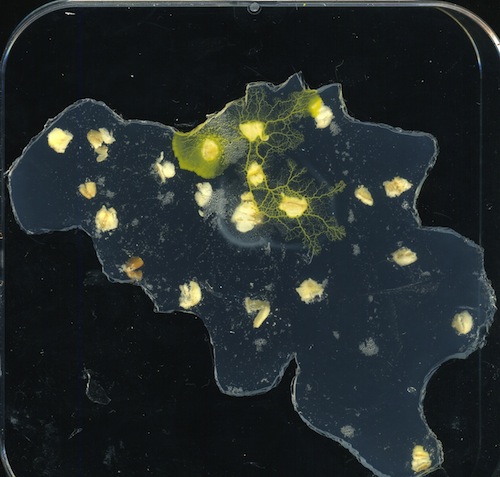}}
\subfigure[24~h]{\includegraphics[width=0.45\textwidth]{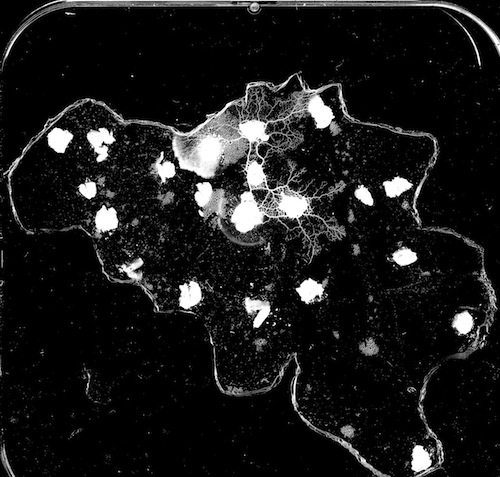}}
\subfigure[48~h]{\includegraphics[width=0.45\textwidth]{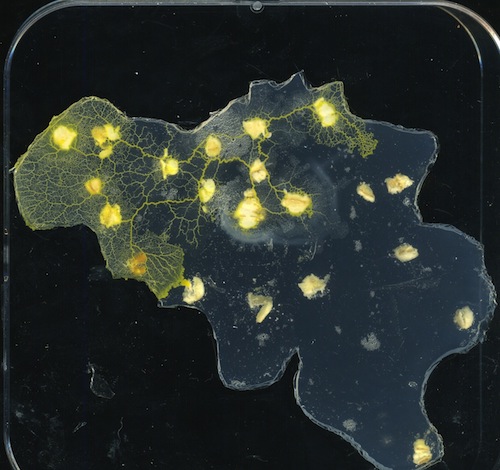}}
\subfigure[48~h]{\includegraphics[width=0.45\textwidth]{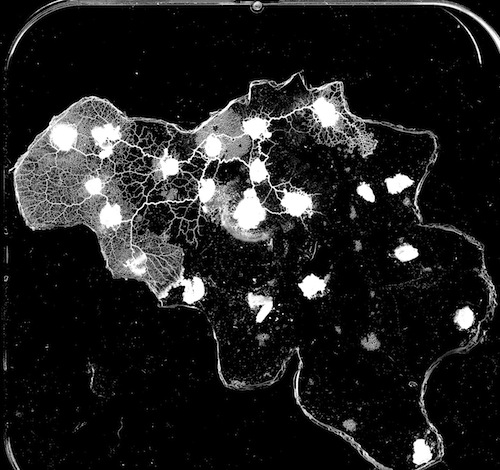}}
\subfigure[72~h]{\includegraphics[width=0.45\textwidth]{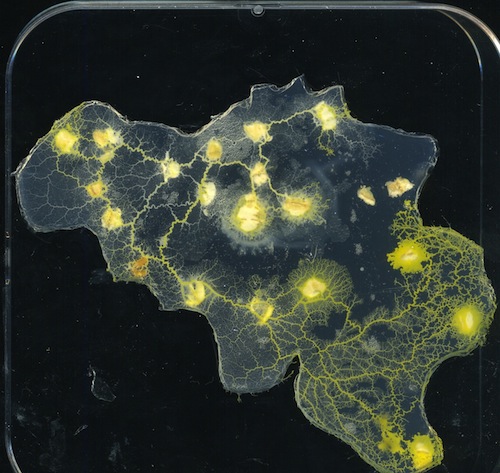}}
\subfigure[72~h]{\includegraphics[width=0.45\textwidth]{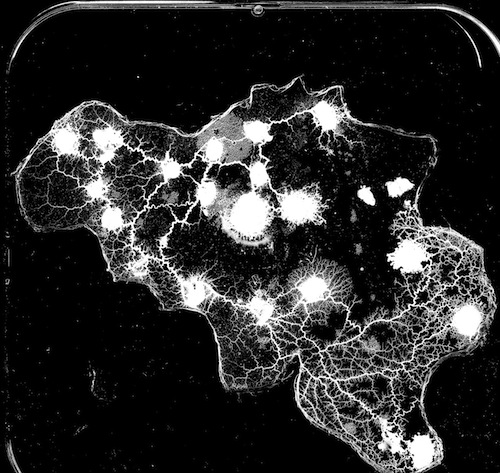}}
\caption{Example of anti-clockwise colonisation, where south of Belgium is colonised via 
Tournai$\rightarrow$Mons$\rightarrow$Charleroi$\rightarrow$Namur route. (ace)~colour scans of the experimental Petri dishes, (bdf)~colour enhanced and grey-scaled versions. Time lapsed after slime mould inoculation in Brussels is shown in captions to sub-figures.}
\label{exampleA11}
\end{figure}

Plasmodium is inoculated in the Brussels area. In the first 24~h it propagates towards and occupies Leuven and Mechelen, 
and then propagates from Mechelen to Antwerp, from Antwerp to Sint-Niklaas (Fig.~\ref{exampleA11}ab).  In the next 24~h slime mould propagates 
from Sint-Niklaas to Aalst and Gent, from Gent to Brugge and from Aalst to Kortrijk. Links from Brugge
to Oostende and Roeselare are built during the same time interval (Fig.~\ref{exampleA11}cd).  Westward development of 
plasmodium is somehow stopped. Despite an attempted propagation from Turnhout towards the Hasselt and Genk areas the slime 
mould never actually reaches these areas in the first 48~h (Fig.~\ref{exampleA11}cd).

\begin{figure}[!tbp]
\centering
\subfigure[96~h]{\includegraphics[width=0.45\textwidth]{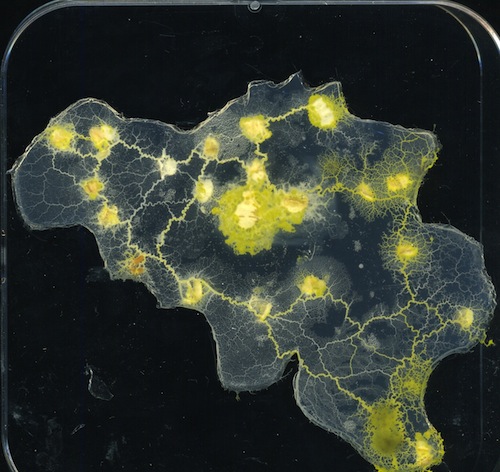}}
\subfigure[96~h]{\includegraphics[width=0.45\textwidth]{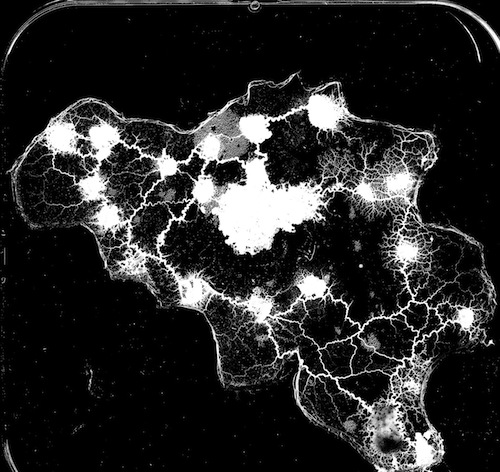}}
\caption{Final stages of anti-clockwise colonisation Fig.~\ref{exampleA11}.}
\label{exampleA11continuation}
\end{figure}

By the 72nd hour after being inoculated in Brussels almost all urban areas but Hasselt and Genk are colonised by slime mould.  Namely in the time interval 48-72~h plasmodium grows from Aalst to Tournai, from Tournai to Mons, and from Mons to 
Charleroi. Slime mould branches at Charleroi and grows in parallel to Namur and Arlon. It propagates from Namur to Sankt-Vith, from Sankt-Vith to Li\`{e}ge. (Fig.~\ref{exampleA11}ef). By the 96th~h after being inoculated the slime mould 
propagates from Li\`{e}ge to Genk and Hasselt areas, and from Hasselt to Leuven. The plasmodium's explorative activities in the Brussels area are 'resumed' when the slime propagates from Leuven and re-occupies the Brussels area (Fig.~\ref{exampleA11continuation} and Fig.~\ref{A11_12_schemes}a.).

\begin{figure}[!tbp]
\centering
\subfigure[48~h]{\includegraphics[width=0.45\textwidth]{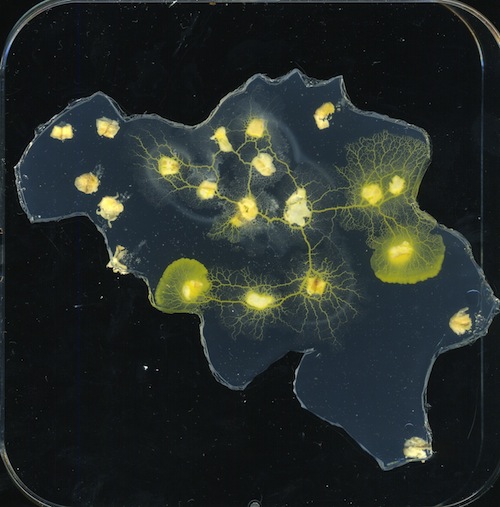}} 
\subfigure[48~h]{\includegraphics[width=0.45\textwidth]{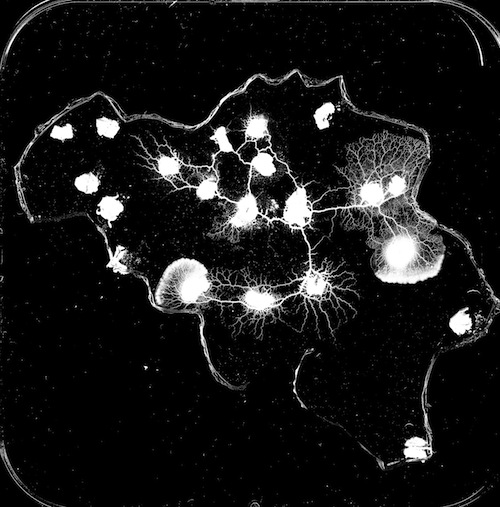}} 
\subfigure[72~h]{\includegraphics[width=0.45\textwidth]{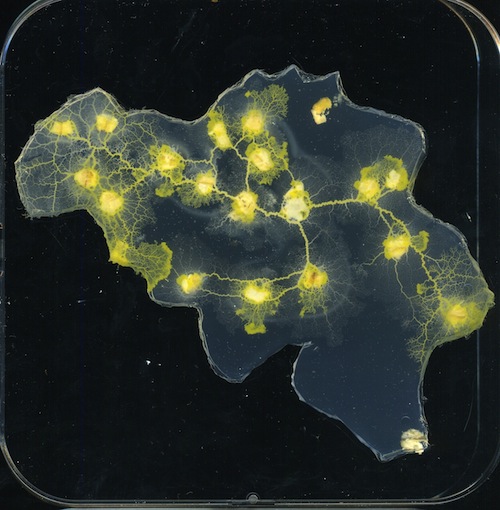}} 
\subfigure[72~h]{\includegraphics[width=0.45\textwidth]{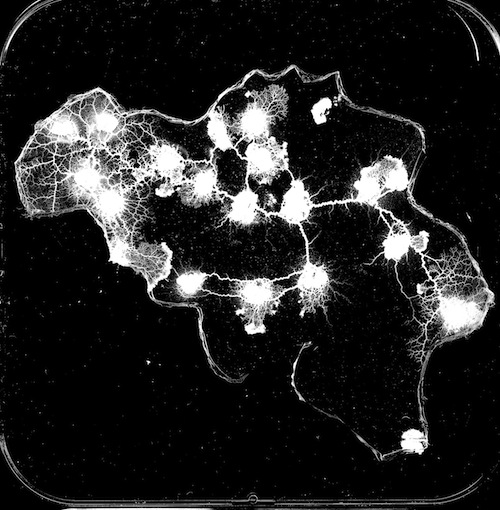}} 
\subfigure[96~h]{\includegraphics[width=0.45\textwidth]{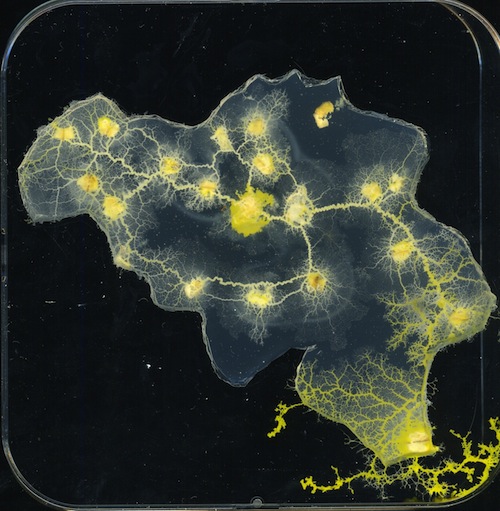}} 
\subfigure[96~h]{\includegraphics[width=0.45\textwidth]{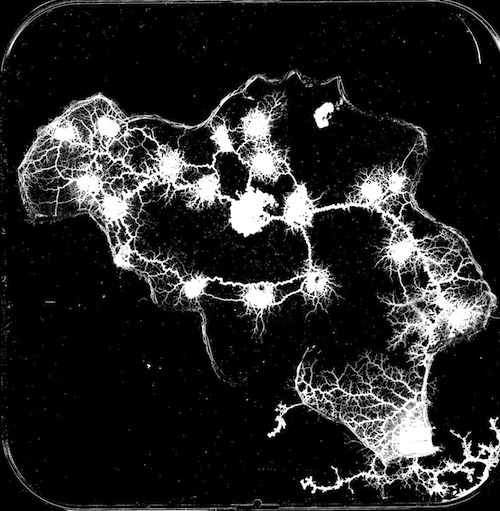}} 
\caption{Example of clockwise colonisation of south of Belgium with dominating route 
Leuven$\rightarrow$Hasselt$\rightarrow$Li\`{e}ge.  (ace)~colour scans of the experimental Petri dishes, (bdf)~colour enhanced and grey-scaled versions.}
\label{exampleA12}
\end{figure}


\begin{figure}[!tbp]
\centering
\subfigure[]{\includegraphics[width=0.49\textwidth]{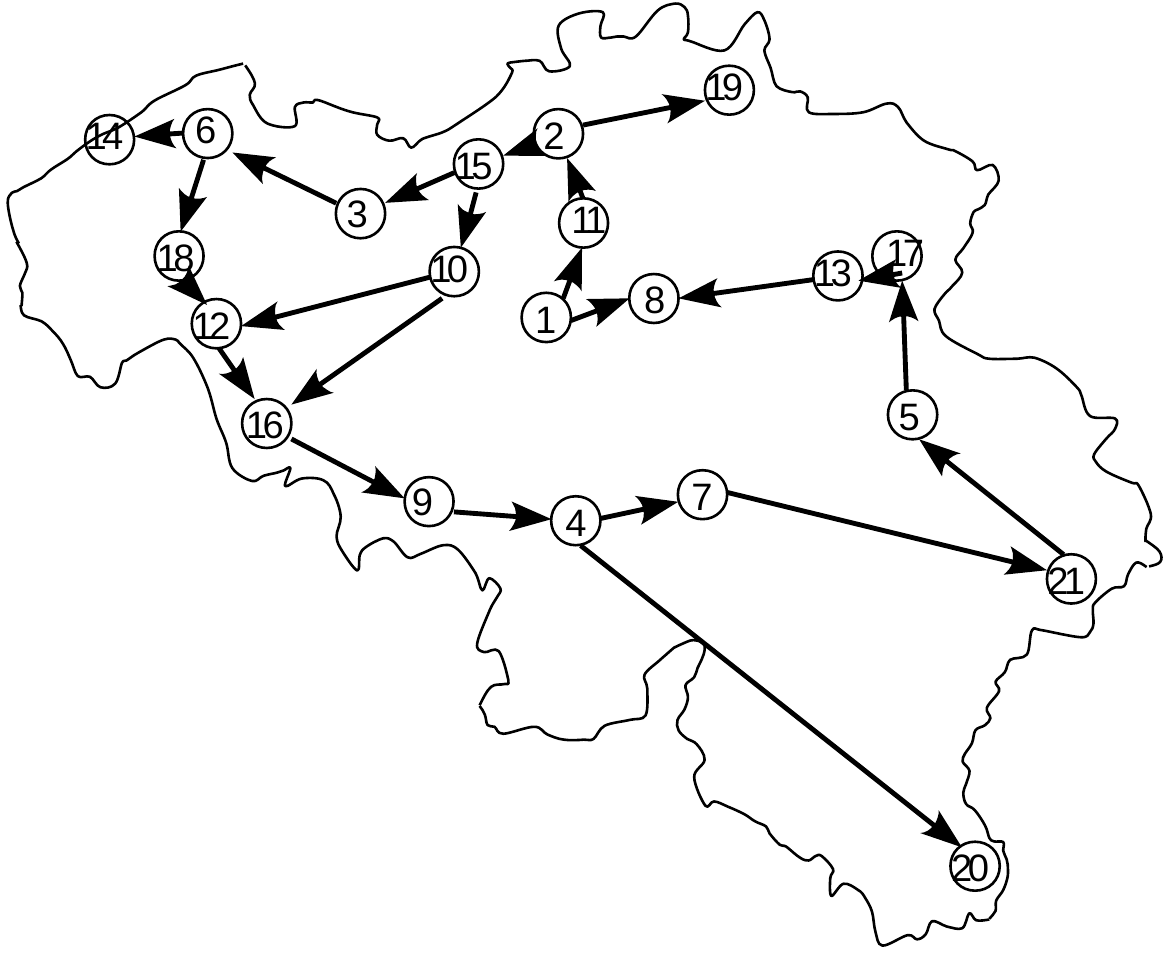}} 
\subfigure[]{\includegraphics[width=0.49\textwidth]{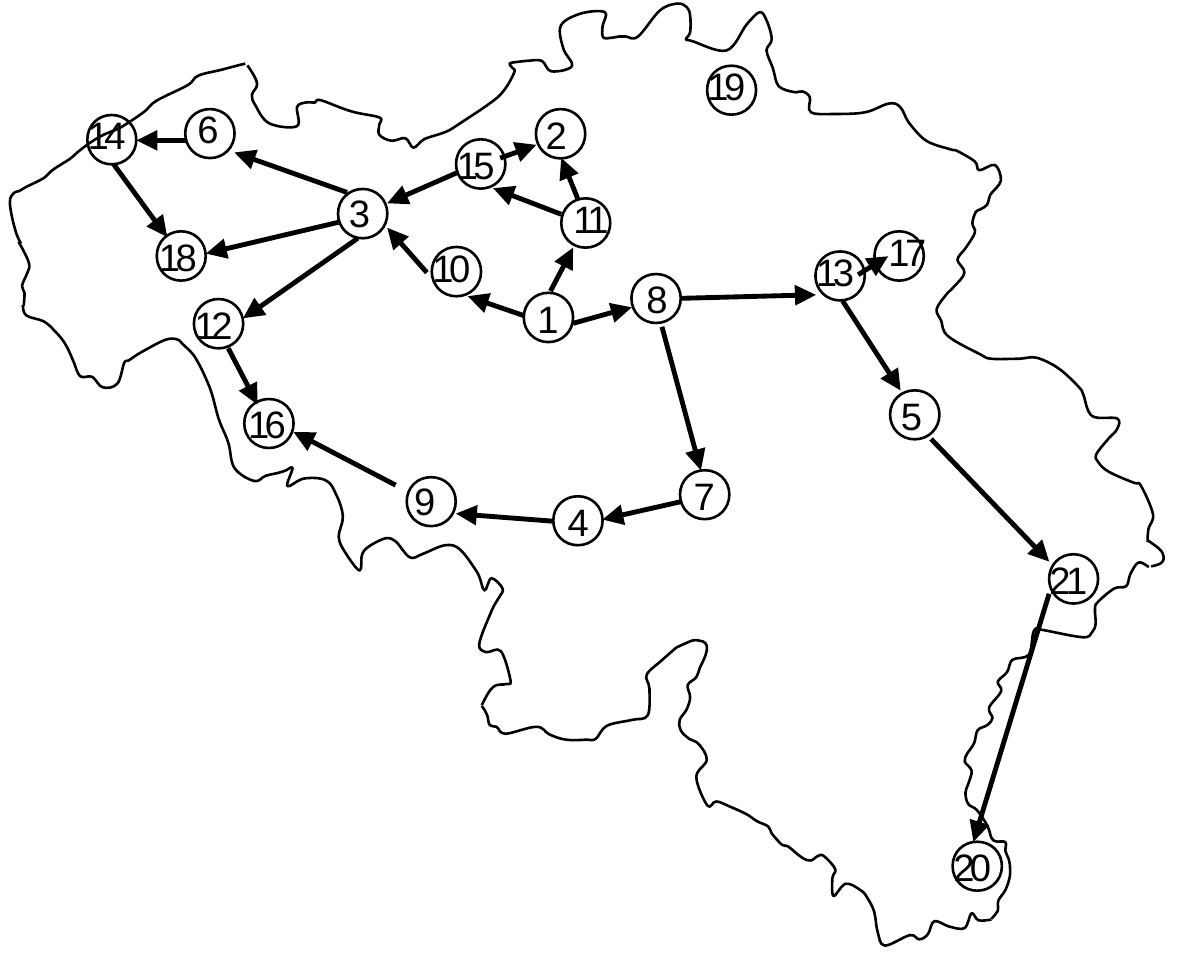}} 
\caption{Scheme of colonisation dynamics illustrated in (a)~Fig.~\ref{exampleA11} and 
(b)~Fig.~\ref{exampleA12}.}
\label{A11_12_schemes}
\end{figure}

Colonisation of Belgium by slime mould shown in Fig.~\ref{exampleA12} develops 
initially according to the scenario described above with the following deviations. Plasmodium propagates from 
Gent to Brugge, Roeselare and Kortrijk at the same time (Figs.~\ref{exampleA12}a--d and~\ref{A11_12_schemes}b). Slime mould grows from 
Leuven to Namur and then to Charleroi. From Charleroi plasmodium propagates to Mons and from Mons to Tournai  (Figs.~\ref{exampleA12}ab). Arlon and Sinkt-Vith are reached by slime mould via Leuven, Hasselt and Li\`{e}ge
(Figs.~\ref{exampleA12}cd and \ref{exampleA12}ef). Antwerp is never colonised by slime mould in this particular experiment (Fig.~\ref{A11_12_schemes}).



\begin{figure}[!tbp]
\centering
\subfigure[$\theta=\frac{1}{28}$]{\includegraphics[width=0.45\textwidth]{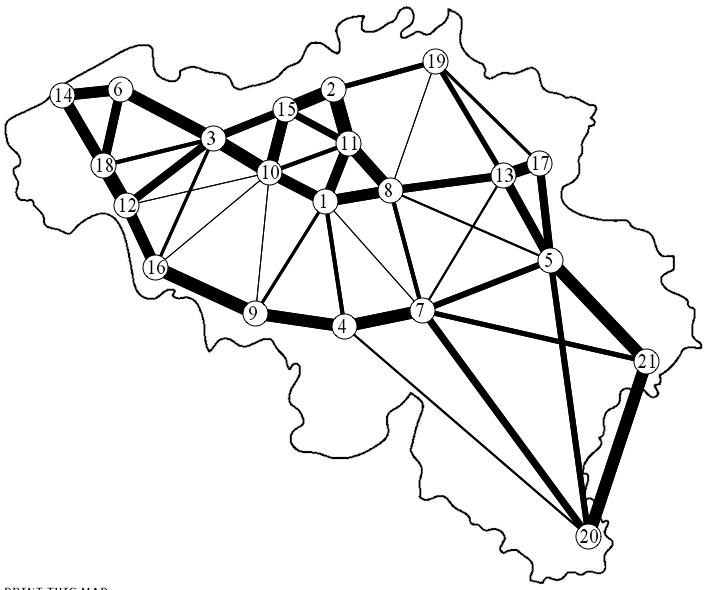}}
\subfigure[$\theta=\frac{6}{28}$]{\includegraphics[width=0.45\textwidth]{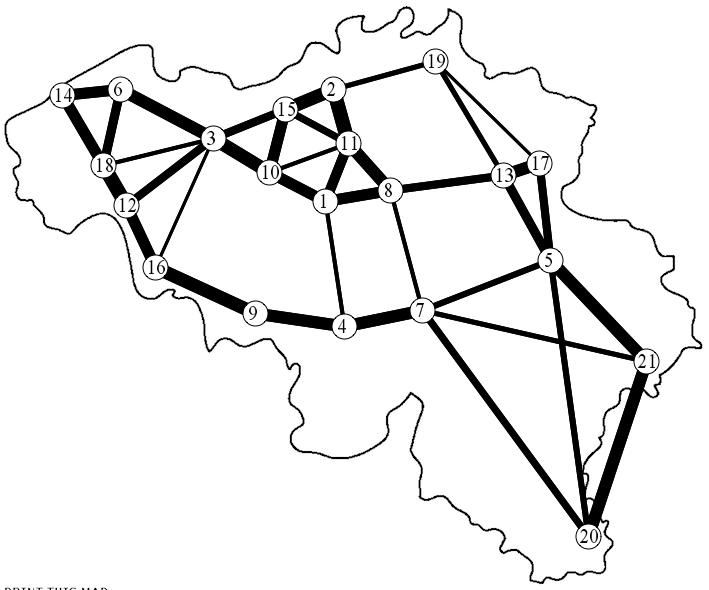}}
\subfigure[$\theta=\frac{10}{28}$]{\includegraphics[width=0.45\textwidth]{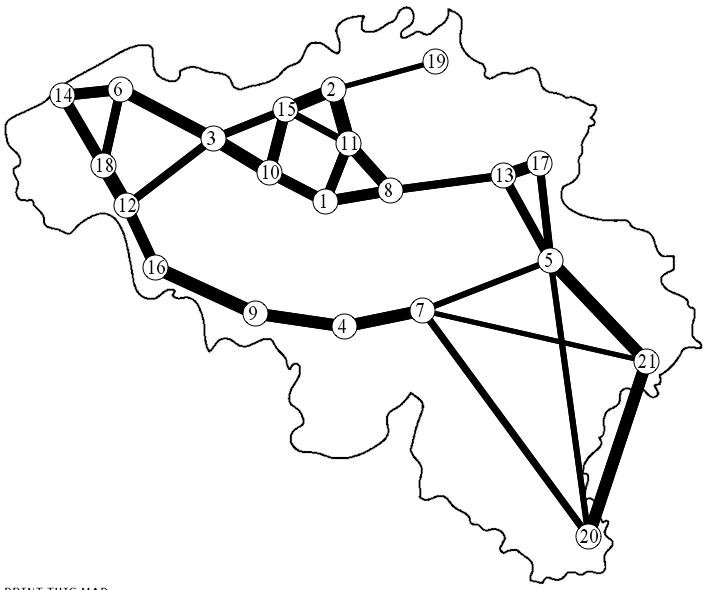}}
\subfigure[$\theta=\frac{11}{28}$]{\includegraphics[width=0.45\textwidth]{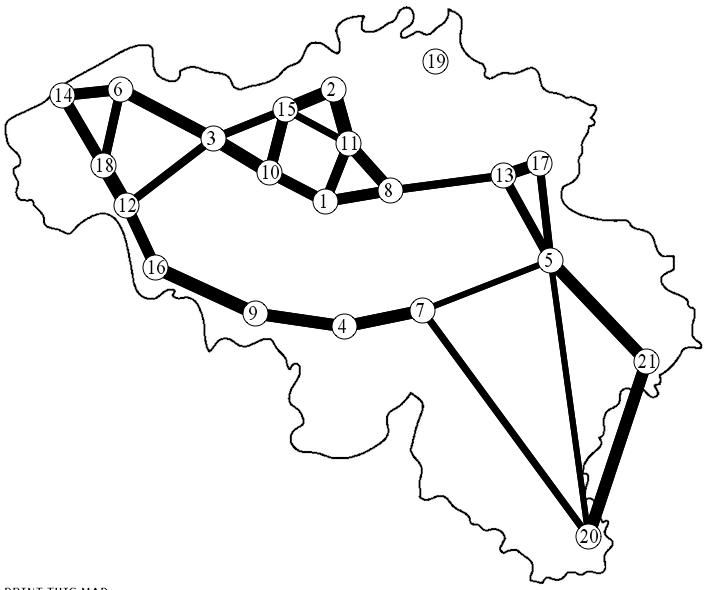}}
\subfigure[$\theta=\frac{16}{28}$]{\includegraphics[width=0.45\textwidth]{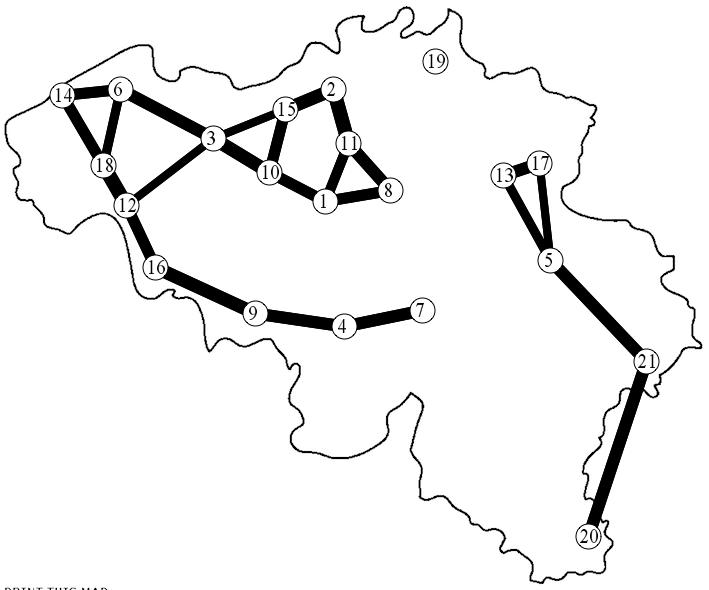}}
\subfigure[$\theta=\frac{22}{28}$]{\includegraphics[width=0.45\textwidth]{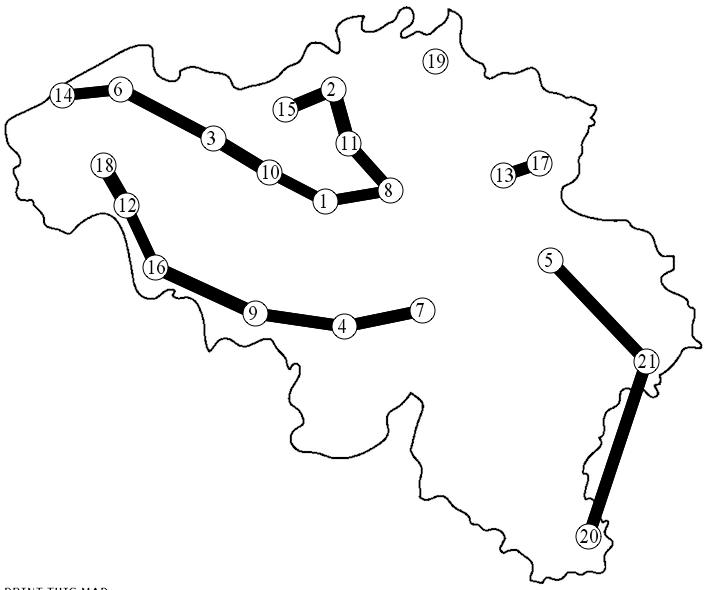}}
\caption{Physarum graphs $\mathbf{P}(\theta)$ for critical values of $\theta$.}
\label{selectedphysarumgraphs}
\end{figure}

\begin{figure}[!tbp]
\centering
$\frac{\includegraphics[width=0.23\textwidth]{figs/PhysarumGraphs/physarum_01}}{\theta=\frac{1}{28}}$
$\frac{\includegraphics[width=0.23\textwidth]{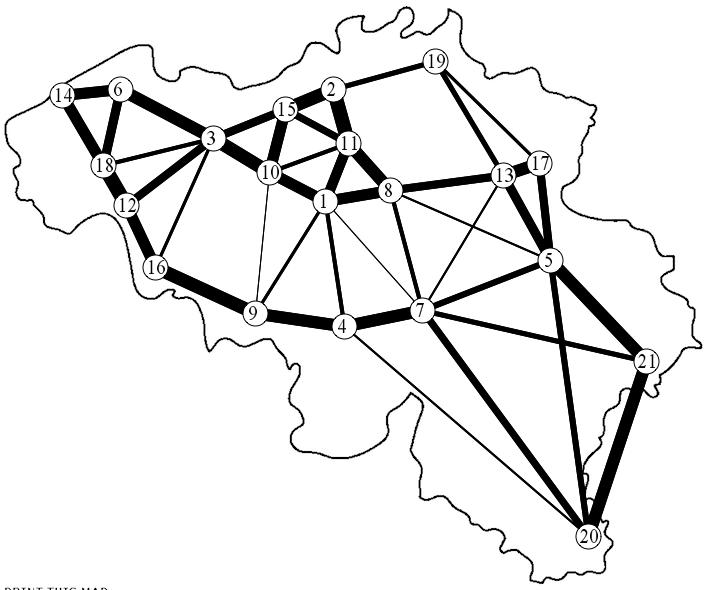}}{\theta=\frac{2}{28}}$
$\frac{\includegraphics[width=0.23\textwidth]{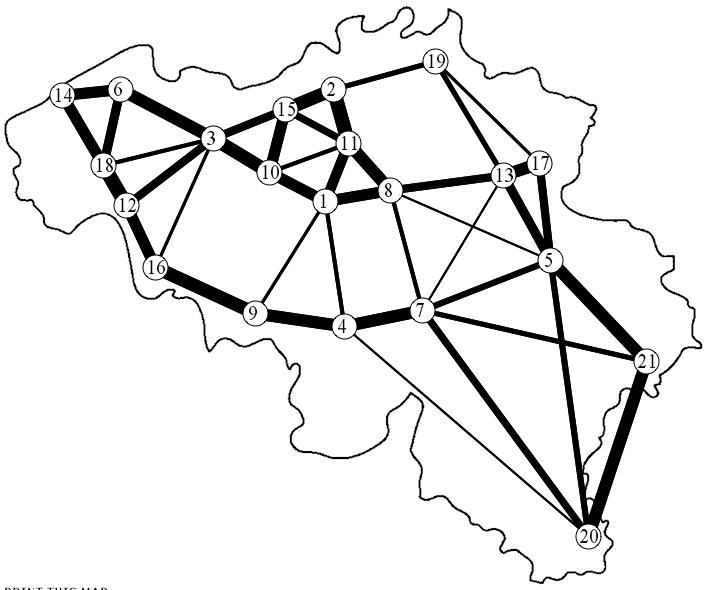}}{\theta=\frac{3}{28}}$
$\frac{\includegraphics[width=0.23\textwidth]{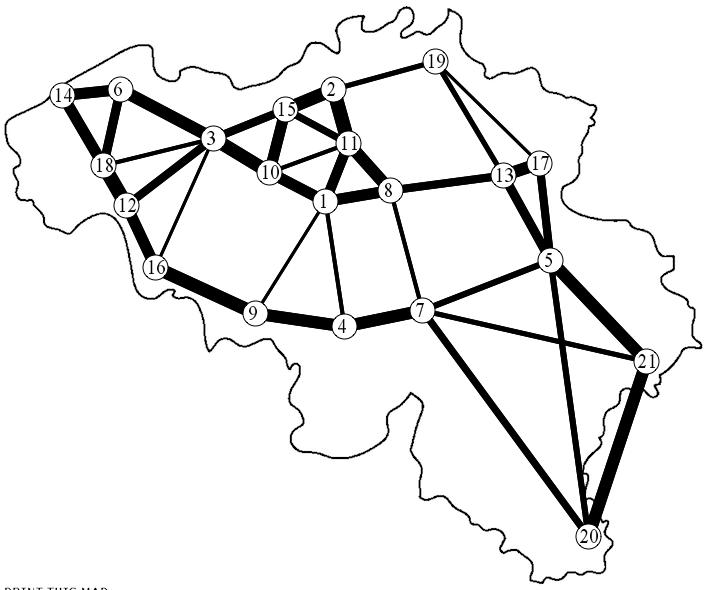}}{\theta=\frac{4}{28}}$
$\frac{\includegraphics[width=0.23\textwidth]{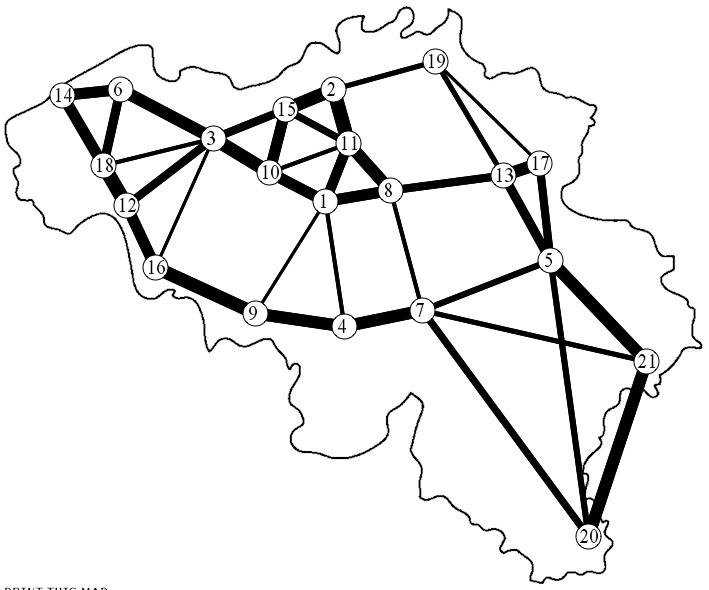}}{\theta=\frac{5}{28}}$
$\frac{\includegraphics[width=0.23\textwidth]{figs/PhysarumGraphs/physarum_06}}{\theta=\frac{6}{28}}$
$\frac{\includegraphics[width=0.23\textwidth]{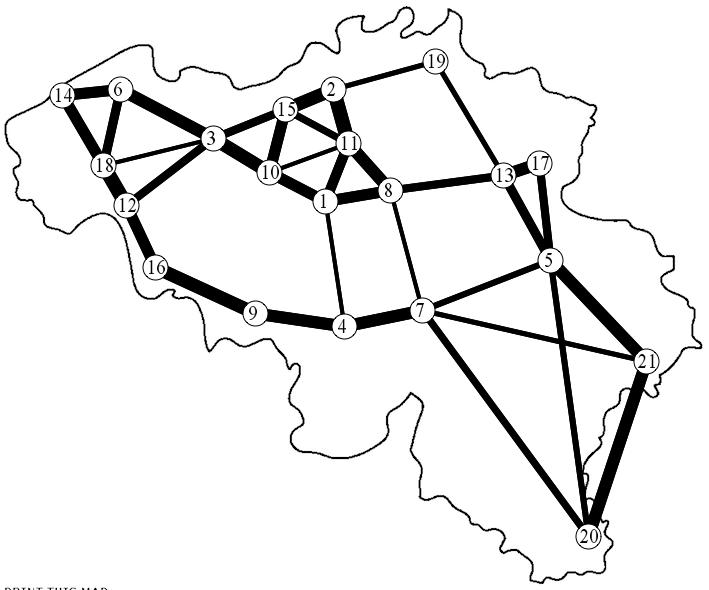}}{\theta=\frac{7}{28}}$
$\frac{\includegraphics[width=0.23\textwidth]{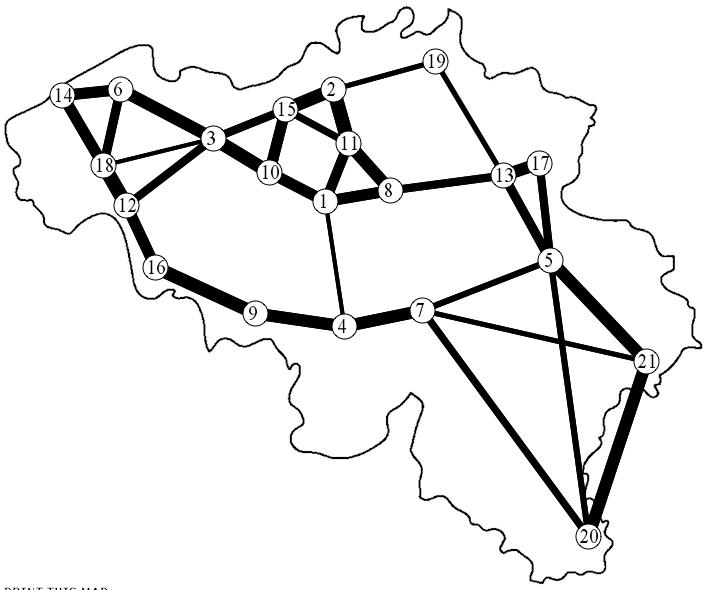}}{\theta=\frac{8}{28}}$
$\frac{\includegraphics[width=0.23\textwidth]{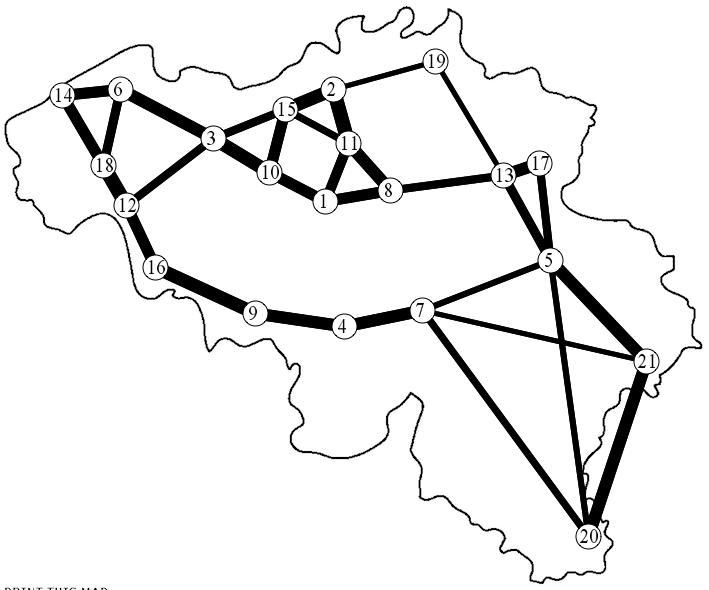}}{\theta=\frac{9}{28}}$
$\frac{\includegraphics[width=0.23\textwidth]{figs/PhysarumGraphs/physarum_10}}{\theta=\frac{10}{28}}$
$\frac{\includegraphics[width=0.23\textwidth]{figs/PhysarumGraphs/physarum_11}}{\theta=\frac{11}{28}}$
$\frac{\includegraphics[width=0.23\textwidth]{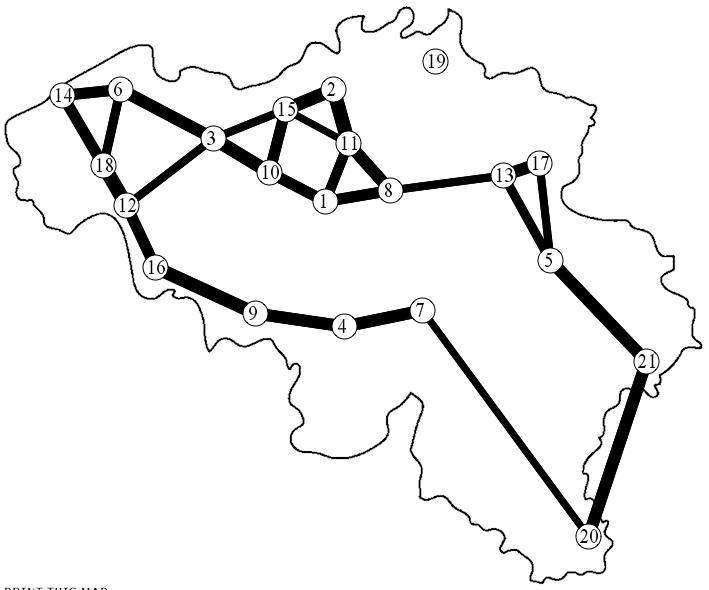}}{\theta=\frac{12}{28}}$
$\frac{\includegraphics[width=0.23\textwidth]{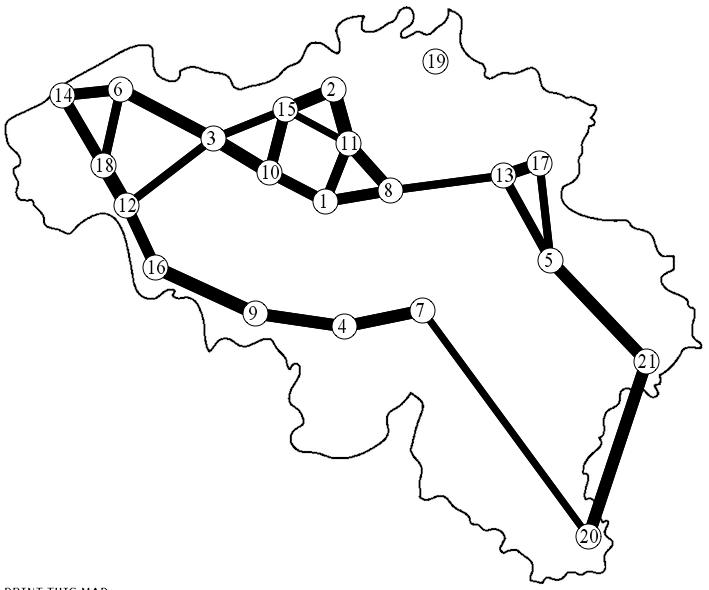}}{\theta=\frac{13}{28}}$
$\frac{\includegraphics[width=0.23\textwidth]{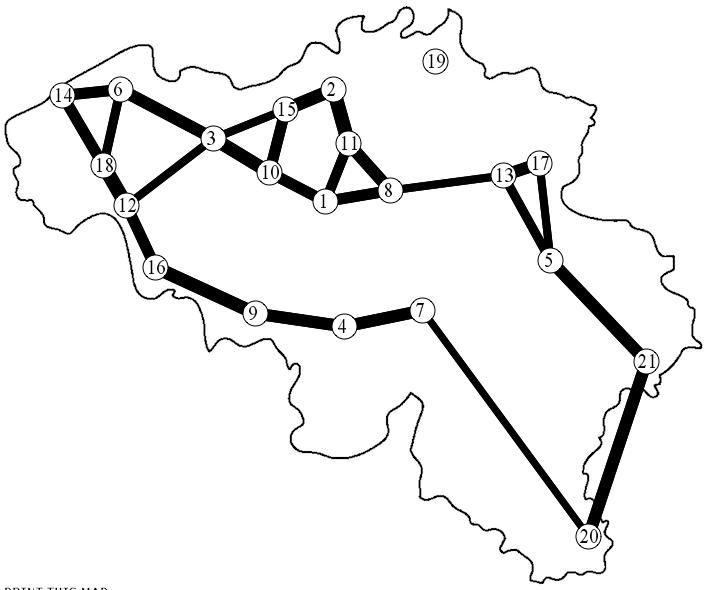}}{\theta=\frac{14}{28}}$
$\frac{\includegraphics[width=0.23\textwidth]{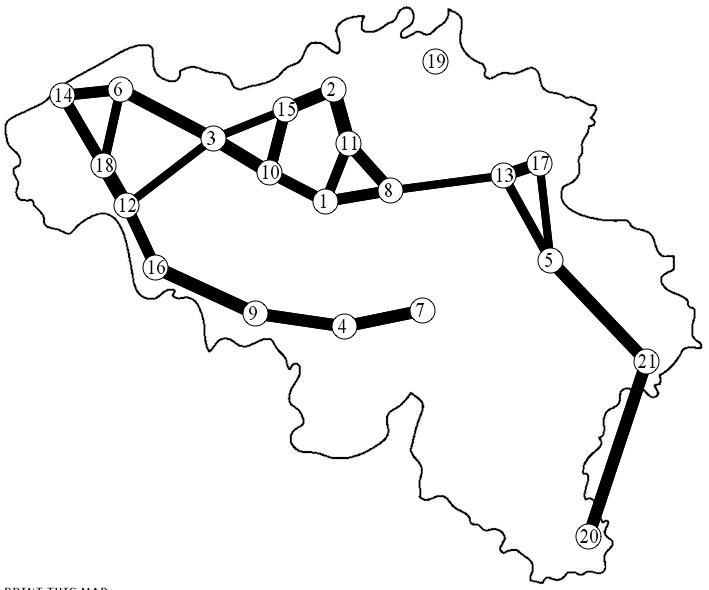}}{\theta=\frac{15}{28}}$
$\frac{\includegraphics[width=0.23\textwidth]{figs/PhysarumGraphs/physarum_16}}{\theta=\frac{16}{28}}$
$\frac{\includegraphics[width=0.23\textwidth]{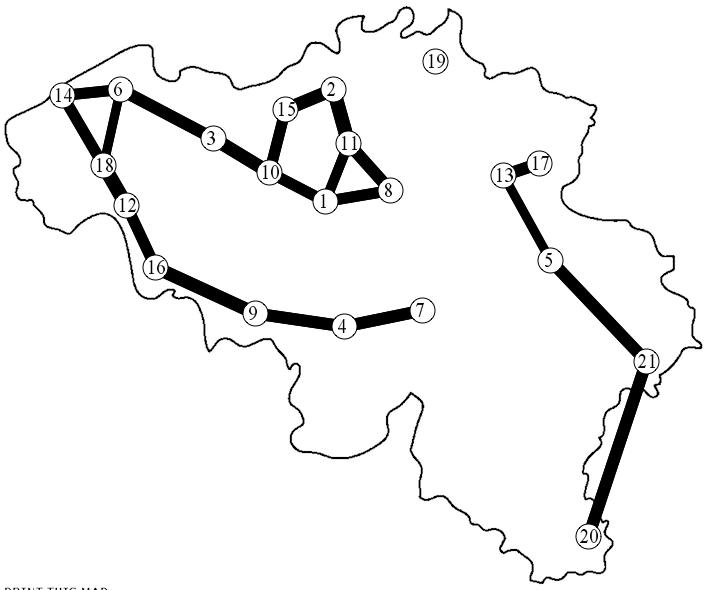}}{\theta=\frac{17}{28}}$
$\frac{\includegraphics[width=0.23\textwidth]{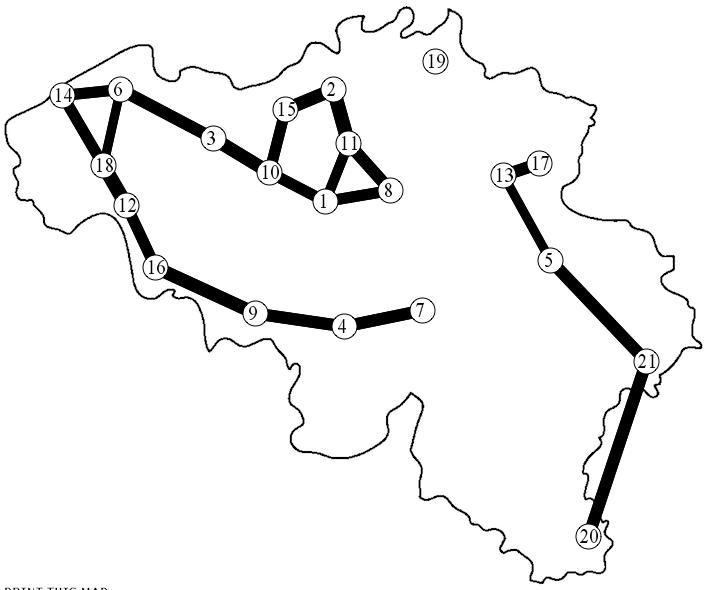}}{\theta=\frac{18}{28}}$
$\frac{\includegraphics[width=0.23\textwidth]{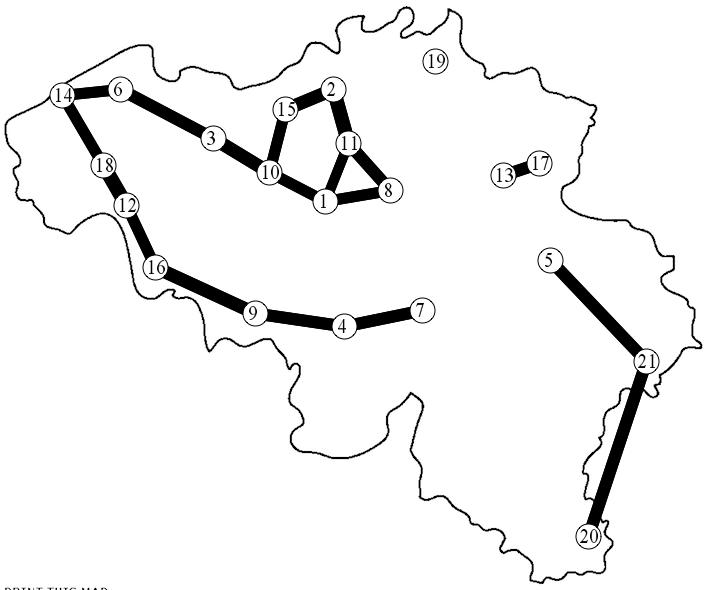}}{\theta=\frac{19}{28}}$
$\frac{\includegraphics[width=0.23\textwidth]{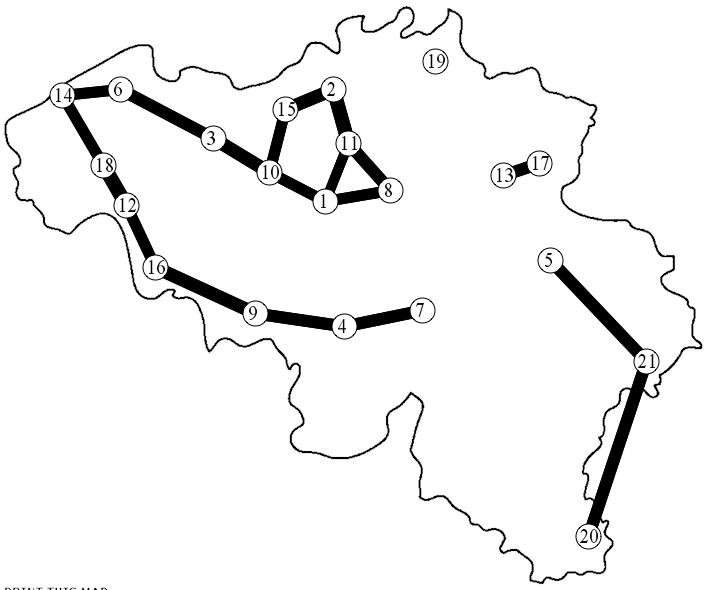}}{\theta=\frac{20}{28}}$
$\frac{\includegraphics[width=0.23\textwidth]{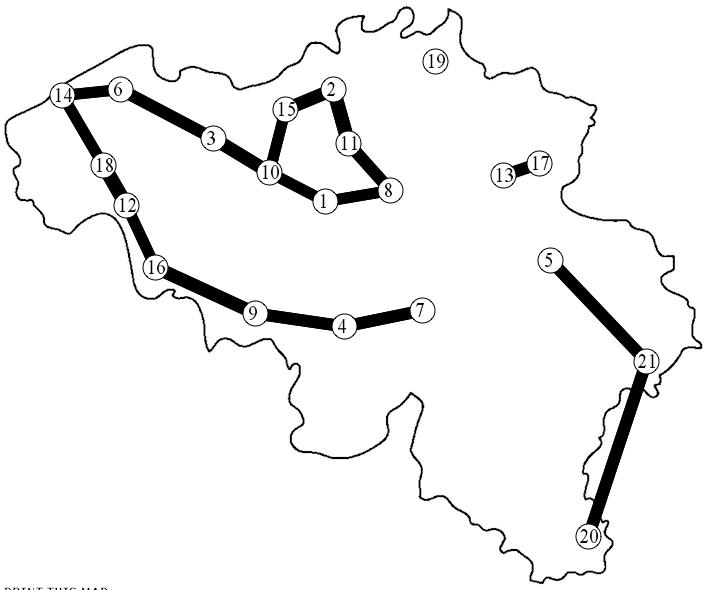}}{\theta=\frac{21}{28}}$
$\frac{\includegraphics[width=0.23\textwidth]{figs/PhysarumGraphs/physarum_22}}{\theta=\frac{22}{28}}$
$\frac{\includegraphics[width=0.23\textwidth]{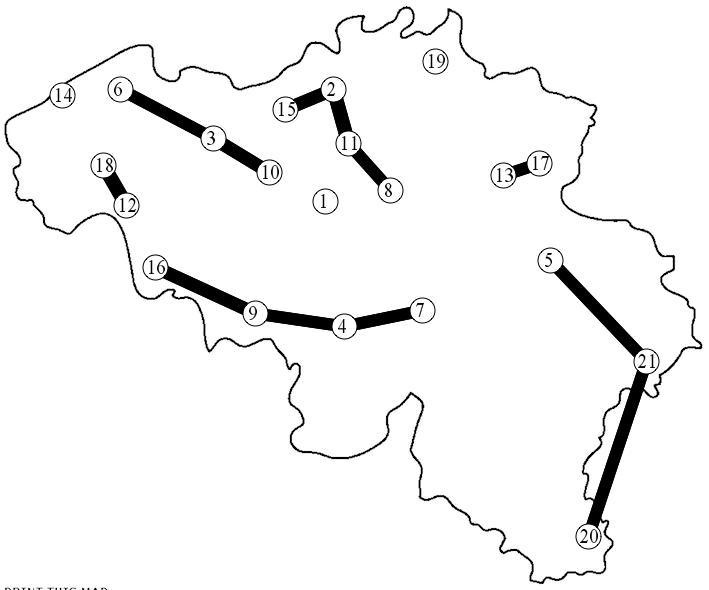}}{\theta=\frac{23}{28}}$
$\frac{\includegraphics[width=0.23\textwidth]{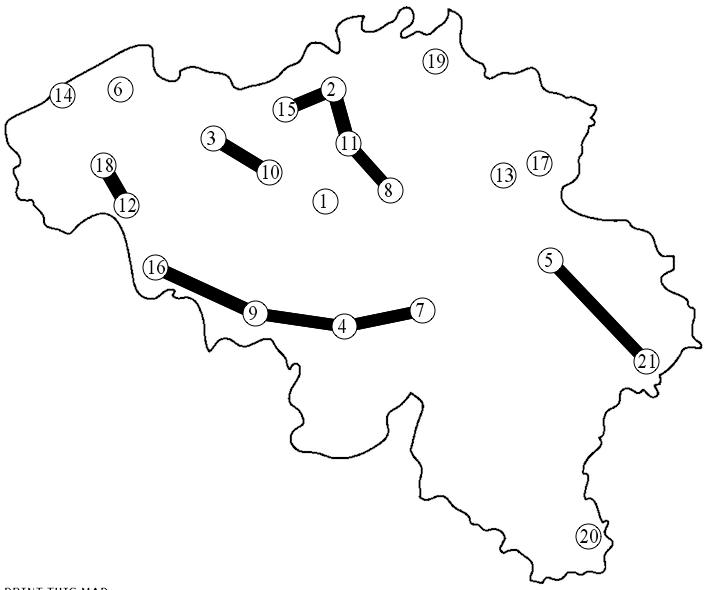}}{\theta=\frac{24}{28}}$
$\frac{\includegraphics[width=0.23\textwidth]{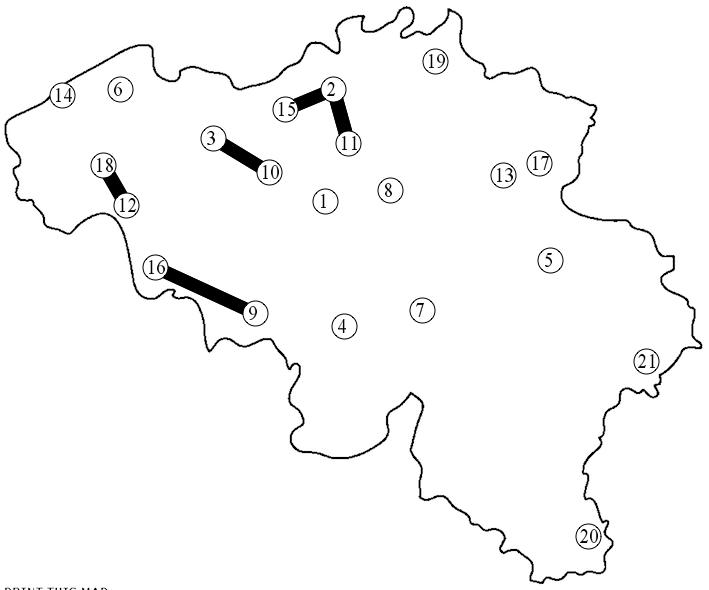}}{\theta=\frac{25}{28}}$
$\frac{\includegraphics[width=0.23\textwidth]{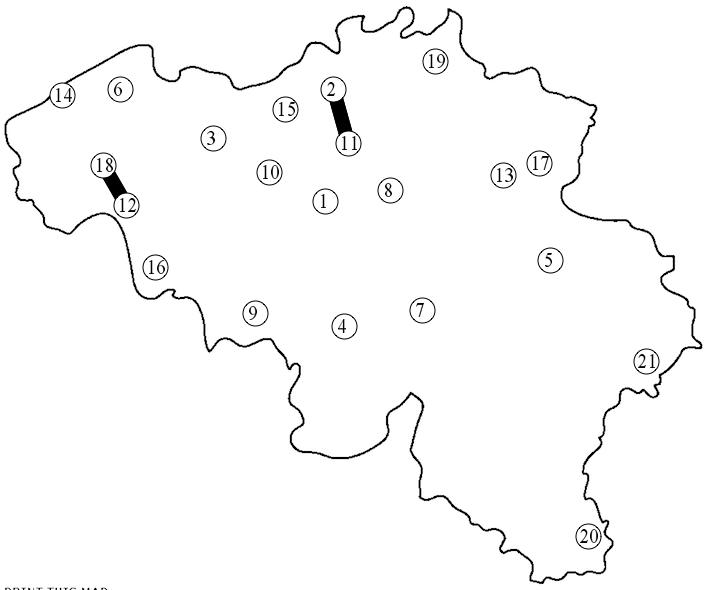}}{\theta=\frac{26}{28}}$
$\frac{\includegraphics[width=0.23\textwidth]{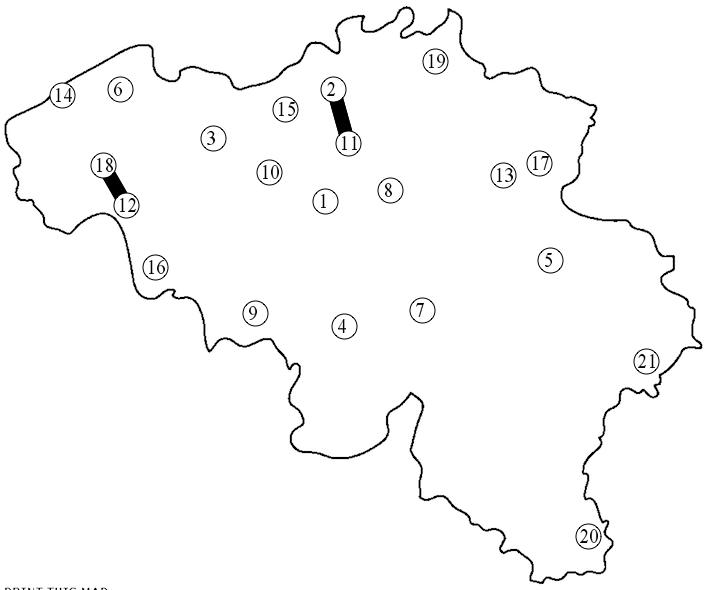}}{\theta=\frac{27}{28}}$
$\frac{\includegraphics[width=0.23\textwidth]{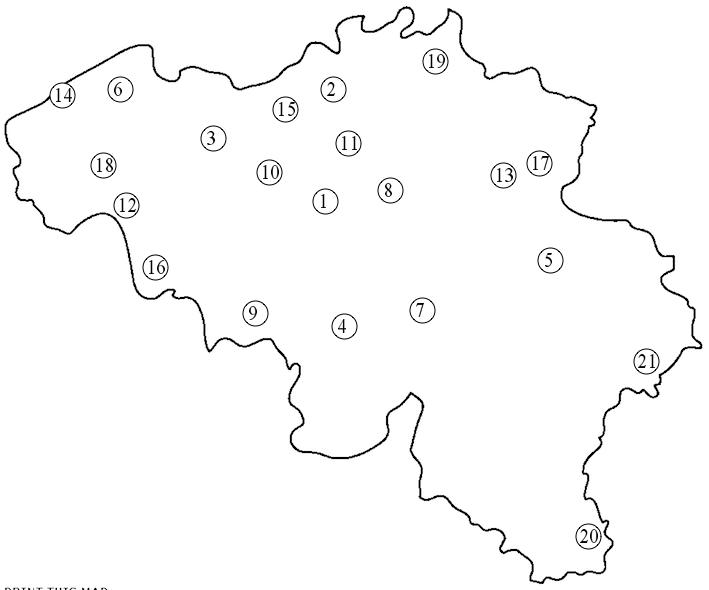}}{\theta=\frac{28}{28}}$
\caption{Generalized Physarum graphs $\mathbf{P}(\theta)$ for $\theta=\frac{1}{28}, \ldots, 1$.}
\label{allphysarumgraphs}
\end{figure}

Two examples discussed above show that the dynamics of colonisation varies from experiment to experiment. 
Therefore from experimental data we extract a generalised Physarum graph. To generalise our experimental results we constructed a Physarum graph with weighted edges. A Physarum graph  is a tuple ${\mathbf P} = \langle {\mathbf U}, {\mathbf E}, w  \rangle$, where $\mathbf U$ is a set of  urban areas, $\mathbf E$ is a set of edges, and
$w: {\mathbf E} \rightarrow [0,1]$ associates to each edge of $\mathbf{E}$   a probability (or weights).
For every two regions $a$ and $b$ from $\mathbf U$ there is an edge connecting $a$ and $b$ if a plasmodium's protoplasmic link is recorded at least in one of $k$ experiments, and the edge $(a,b)$ has a probability calculated as the ratio of experiments where protoplasmic link $(a,b)$ occurred in the total number of experiments ($k=23$). For example, 
if we observed a protoplasmic tube connecting areas $a$ and $b$ in 7 experiments, the weight of edge $(a,b)$ will be $w(a,b)=\frac{7}{23}$. We do not take into account the exact configuration of the protoplasmic tubes but merely their existence. Further we will be dealing with threshold Physarum graphs $\mathbf{P}(\theta)  = \langle  {\mathbf U}, T({\mathbf E}), w, \theta \rangle$. A threshold Physarum graph is obtained from Physarum graph by the transformation: $T({\mathbf E})=\{ e \in {\mathbf E}: w(e) geq \theta \}$. That is, all edges with weight less than  $\theta$ are removed.  Examples of threshold Physarum graphs for various values of $\theta$ are shown in 
Fig.~\ref{allphysarumgraphs}.

The 'raw' Physarum graph $\mathbf{P}(\frac{1}{28})$ is a non-planar\footnote{A planar graph consists of nodes which are points of the  Euclidean plane and edges which are straight segments connecting the points.} acyclic graph (Fig.~\ref{selectedphysarumgraphs}a). In a raw graph each edge appears at least in one laboratory experiment. We call an edge of Physarum  
\emph{credible} if this edge is represented by a protoplasmic tube in over 20\% of laboratory experiments. Such graph  $\mathbf{P}(\frac{6}{28})$ is shown in Fig.~\ref{selectedphysarumgraphs}b. It is still non-planar, however, the only intersecting edges are the links Li\`{e}ge area --- Arlon and Namur --- Sankt-Vith. 

\begin{finding}
The Physarum graph with all credible edges is a non-planar cyclic graph. 
\end{finding}

The graph remains connected while $\theta$ grows up to $\frac{10}{28}$ (Fig.~\ref{selectedphysarumgraphs}c). The Turnhout urban area becomes an isolated vertex when $\theta=\frac{11}{28}$ (Fig.~\ref{selectedphysarumgraphs}d). For this value of $\theta$ the Physarum graph becomes planar because the link (Li\`{e}ge area -- Arlon) is represented in more experiments than the link (Namur -- Sankt-Vith). The graph  $\mathbf{P}(\frac{11}{28})$ has the largest (among all Physarum graphs studied here) empty, i.e. not having any edges inside, circle. Clockwise, starting in the Brussels area it spans Leuven, Hasselt, Li\`{e}ge, Namur, Charleroi, Mons, Tournai, Kortrijk, Gent, Aalst and finishes in Brussels.

The Physarum graph $\mathbf{P}(\theta)$ splits into three components when $\theta=\frac{16}{28}$. The smallest component is the isolated vertex Turnhout. The medium size component is a cycle Li\`{e}ge --- Hasselt --- Genk --- Li\`{e}ge attached to a segment Arlon --- Sank-Vith (Fig.~\ref{selectedphysarumgraphs}e). The largest component is a proximity graph spanning all remaining urban areas.

\begin{finding}
The following slime mould transport chains appear in almost all experiments  (Fig.~\ref{selectedphysarumgraphs}f):
\begin{itemize}
\item chain Roeselare --- Kortrijk --- Mons --- Charleroi --- Namur 
\item chain Oostende --- Brugge ---  Gent ---  Aalst --- Brussels --- Leuven --- Mechelen --- Antwerp --- Sint-Niklaas 
\item chain Li\`{e}ge --- Sankt-Vith --- Arlon
\item one-link chain Hasselt ---  Genk. 
\end{itemize} 
\end{finding}

\section{Slime mould versus Belgian motorways}
\label{comparingPhysarumMotorways}

\begin{figure}[!tbp]
\centering
\subfigure[]{\includegraphics[width=0.8\textwidth]{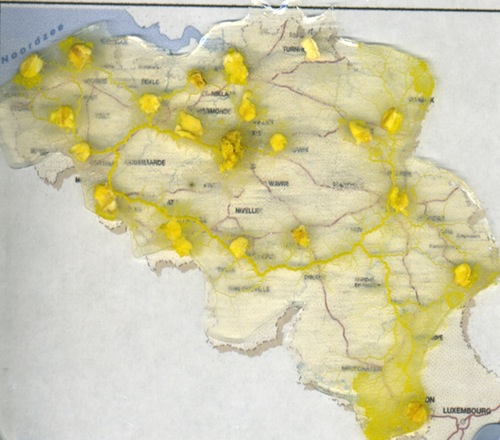}}
\subfigure[]{\includegraphics[width=0.8\textwidth]{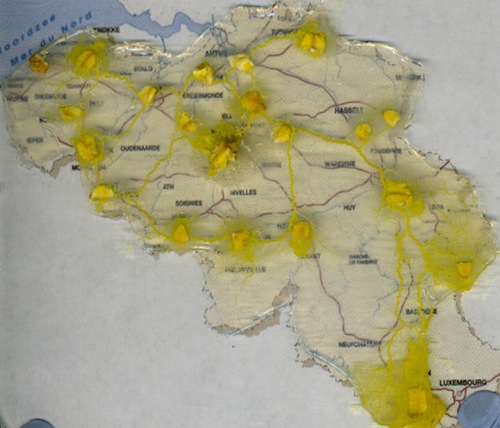}}
\caption{Examples of \emph{P. polycephalum} protoplasmic networks on the Belgian motorway network.}
\label{PhysarumOnMotorwayMap}
\end{figure}

As we can see in the examples of the experimental configurations in Fig.~\ref{PhysarumOnMotorwayMap}, plasmodium networks are polymorphic and no two networks are exactly the same. Some motorways are matched by protoplasmic tubes well, others just approximated, and some do not have a slime mould representation at all.

For example, Fig.~\ref{PhysarumOnMotorwayMap}a demonstrates that slime mould developed protoplasmic tubes corresponding to 
motorway A10/E40 between Gent and Brugge, E17 between Gent and Kortrijk, E17 and E403/A17 between Gent and Tournai, and E42 connecting Mons to Charleroi to Namur to Li\`{e}ge. Motorways E42 (Tournai -- Mons), E46 (Antwerp -- Turnhout) and E40 (Leuven -- Li\`{e}ge) are represented by slime mould in the experiment illustrated in   Fig.~\ref{PhysarumOnMotorwayMap}b. At the same time, the transport link E411 from Brussels to Namur to Arlon is not represented in the configurations in Fig.~\ref{PhysarumOnMotorwayMap}.

Due to the variability of the slime mould networks it would be unreasonable to attempt comparing the exact topology of slime mould and 
man-made transport links. Instead we will compare them at the level of graph-theoretical representations.

\begin{figure}[!tbp]
\centering
\includegraphics[width=0.7\textwidth]{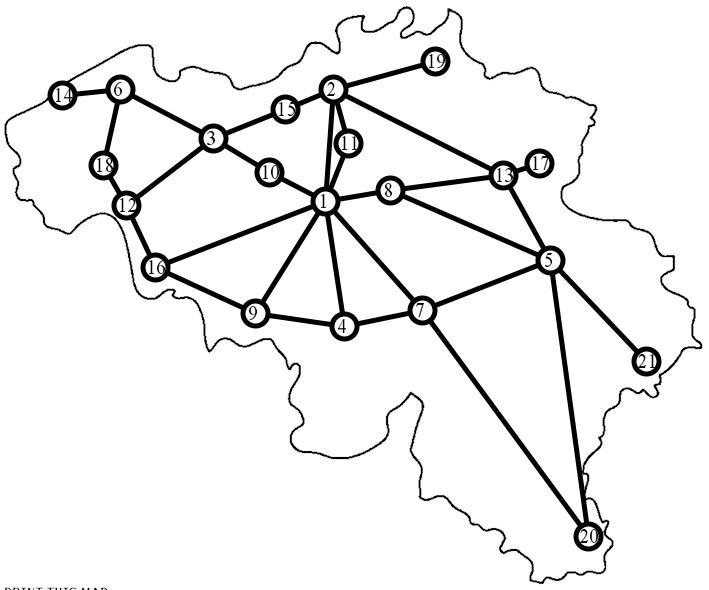}
\caption{Graph $\mathbf H$ of Belgian motorway network.}
\label{motorways}
\end{figure}

The graph $\mathbf H$ of the Belgian motorway network (Fig.~\ref{motorways}) is constructed as follows. Let $\mathbf U$ be a set of urban regions/cities; for any two regions $a$ and $b$ from $\mathbf U$, the nodes $a$ and $b$ are connected by an edge $(a,b)$ if there is a motorway starting in the vicinity of $a$, passing in the vicinity of $b$, and not passing in the vicinity of any other urban area $c \in \mathbf U$. In the case of branching -- that is, a motorway starts in $a$, goes in the direction of $b$ and $c$, and at some point branches towards $b$ and $c$ -- we then add two separate edges $(a,b)$ and $(a,c)$ to the graph $\mathbf H$.  The motorway graph is planar (Fig.~\ref{motorways}).

\begin{figure}[!tbp]
\centering
\subfigure[$\mathbf{H} \cap \mathbf{P}(\frac{1}{28})$]{\includegraphics[width=0.45\textwidth]
{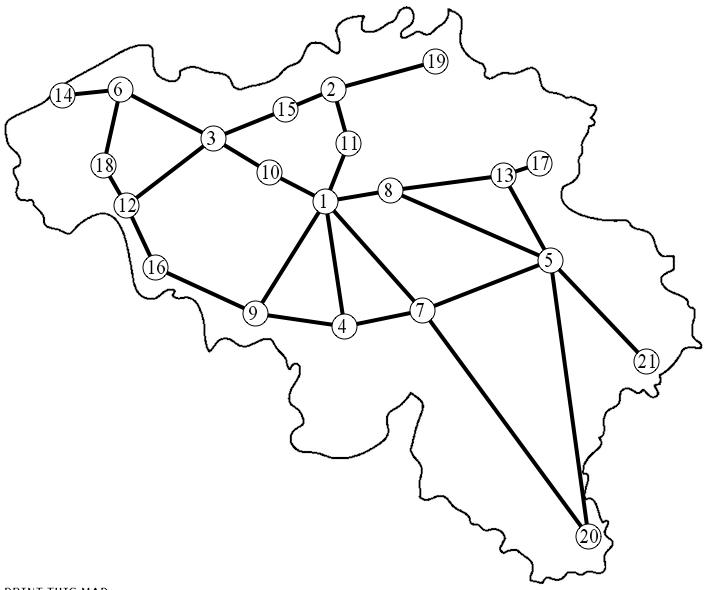}}
\subfigure[$\mathbf{H} \cap \mathbf{P}(\frac{6}{28})$=$\mathbf{H} \cap \mathbf{P}(\frac{10}{28})$]{\includegraphics[width=0.45\textwidth]
{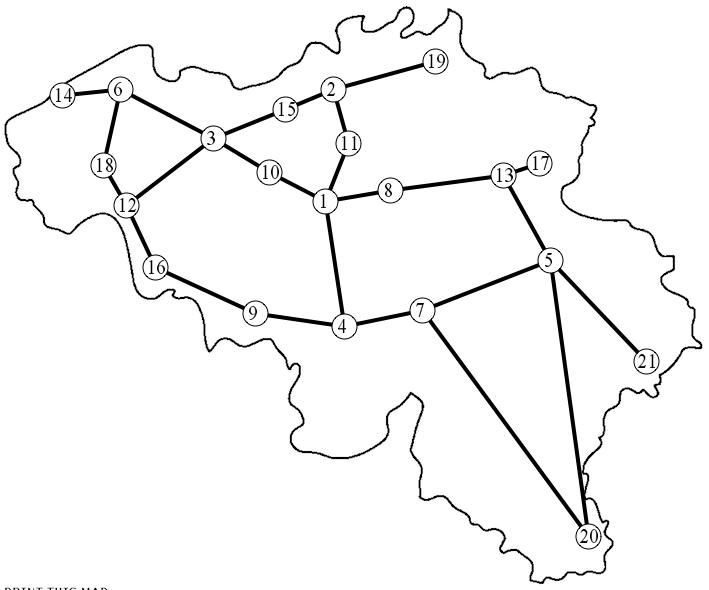}}
\subfigure[$\mathbf{H} \cap \mathbf{P}(\frac{11}{28})$]{\includegraphics[width=0.45\textwidth]
{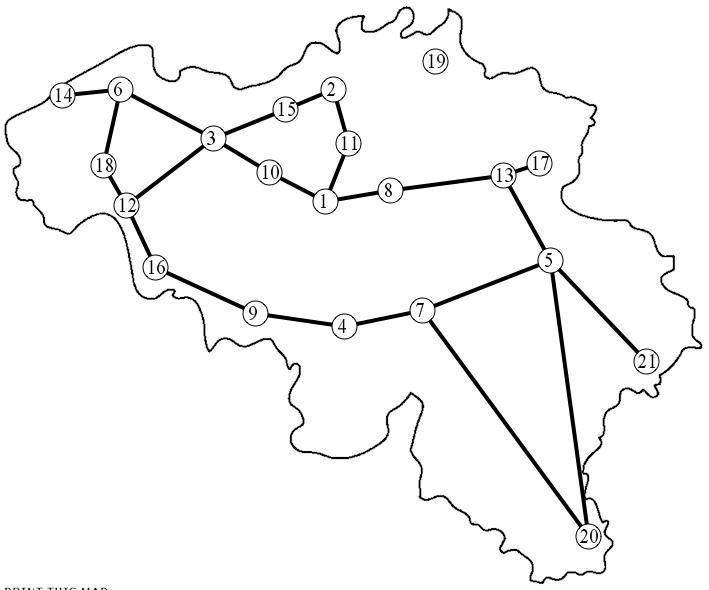}}
\subfigure[$\mathbf{H} \cap \mathbf{P}(\frac{16}{28})$]{\includegraphics[width=0.45\textwidth]
{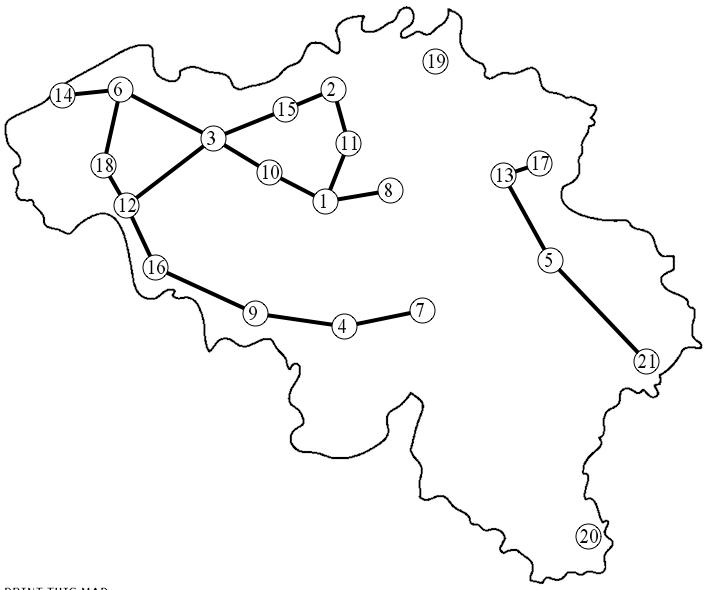}}
\subfigure[$\mathbf{H} \cap \mathbf{P}(\frac{22}{28})$]{\includegraphics[width=0.45\textwidth]
{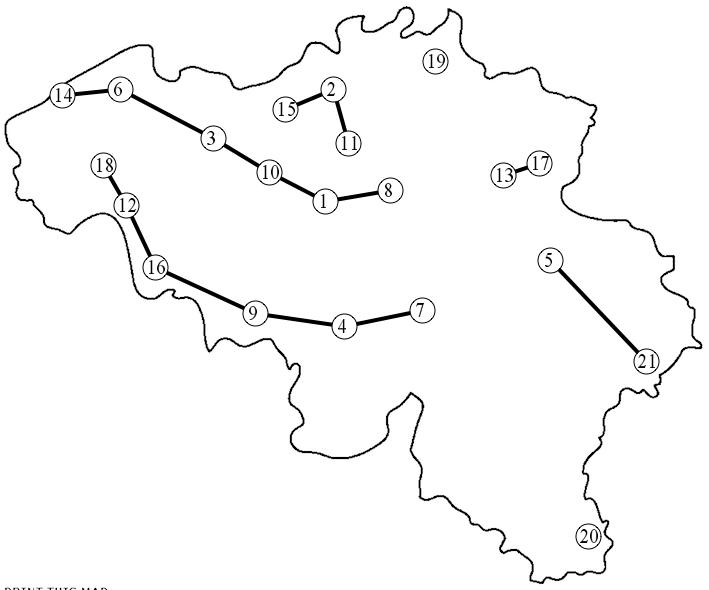}}
\caption{Intersection $\mathbf{H} \cap  \mathbf{P}(\theta)$ of motorway graph $\mathbf{H}$
with Physarum graphs $\mathbf{P}(\theta)$ for critical values of $\theta$.}
\label{IntersectPhysarumMotorways}
\end{figure}

\begin{finding}
Physarum polycephalum almost completely approximates the Belgian motorway network.
\end{finding}

Namely, $\mathbf{H} \nsubseteq \mathbf{P}(\frac{1}{28})$ but 25 of 28 edges of $\mathbf{H}$ 
are edges of  $\mathbf{P}(\frac{1}{28})$. The following motorway links are never represented by 
protoplasmic tubes, not in a single experiment: Brussels to Tournai, Brussels to Antwerp, Antwerp 
to Hasselt (Fig.~\ref{IntersectPhysarumMotorways}a and Fig.~\ref{motorways}).

\begin{finding}
\textbf{On redundancy}. Motorway links connecting Brussels with Antwerp, Tournai, Mons, Charleroi, Namur, and links connecting Leuven with Li\`{e}ge and Antwerp with Genk and Turnhout are proved to be redundant  components of the Belgian transport system
in the slime mould experiments. 
\end{finding}

The skeletal motorway network represented by the plasmodium network in laboratory experiments is shown in 
Fig.~\ref{IntersectPhysarumMotorways}b. This network is represented by protoplasmic tubes in over 35\% 
of experiments; and in almost 40\% of experiments without the Antwerp --- Turnhout link (Fig.~\ref{IntersectPhysarumMotorways}cd). By increasing the level of slime mould's ``confidence"  to almost 60\% we loose motorway links connecting Leuven with Hasselt, Namur with Li\`{e}ge and Arlon, and 
Li\`{e}ge with Arlon (Fig.~\ref{IntersectPhysarumMotorways}cd).

\begin{finding}
Motorway segments A17 (Antwerp --- Sint-Niklaas), A19 (Antwerp --- Mechelen),  E42 (Li\`{e}ge --- Sankt-Vith), 
A10/E40 (Oostende --- Gent --- Aalst --- Brussels --- Leuven), A17/E403 (Roeselare --- Kortrijk --- Tournai), E42/E19 (Tournai --- Mons), E42 (Mons --- Charleroi --- Namur) are essential transport links from the slime mould's point of view. 
\end{finding}

As illustrated in Fig.~\ref{IntersectPhysarumMotorways}f these transport links are represented by the protoplasmic network in almost 80\% of laboratory experiments.

\section{Proximity graphs}
\label{proximity}

\begin{figure}[!tbp]
\centering
\subfigure[${\mathbf{RNG}}$]{\includegraphics[width=0.49\textwidth]{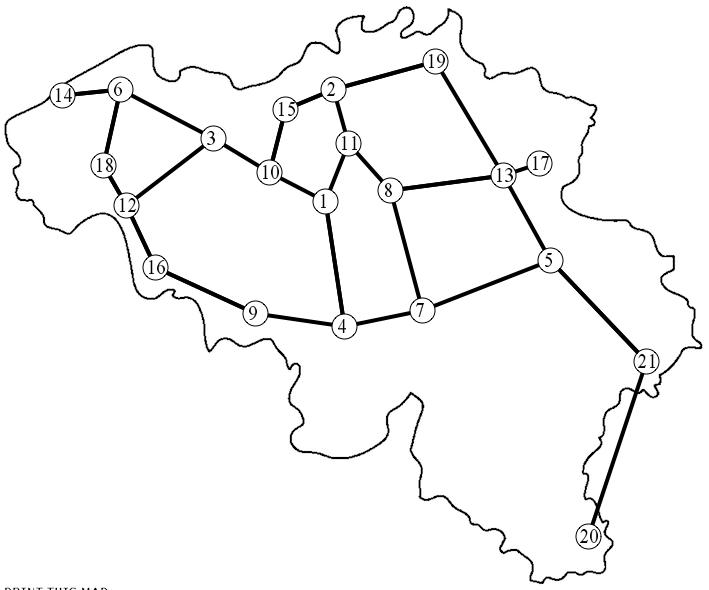}}
\subfigure[${\mathbf{GG}}$]{\includegraphics[width=0.49\textwidth]{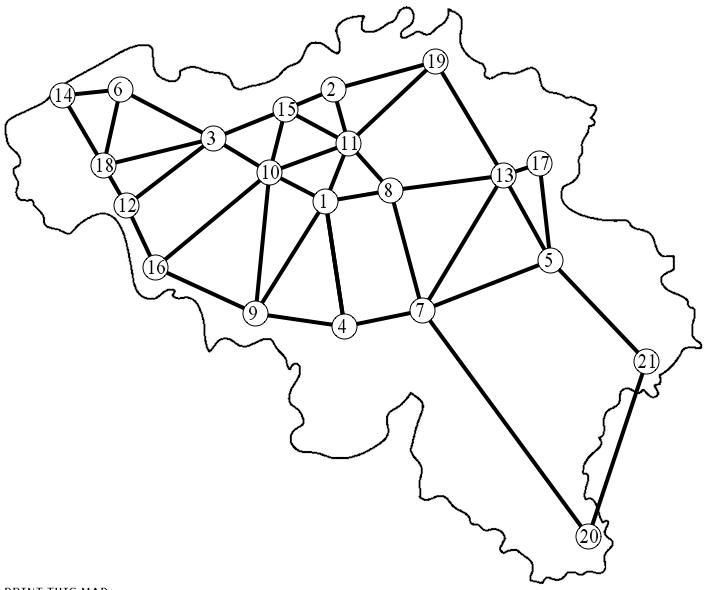}}
\subfigure[${\mathbf{MST}}$]{\includegraphics[width=0.49\textwidth]{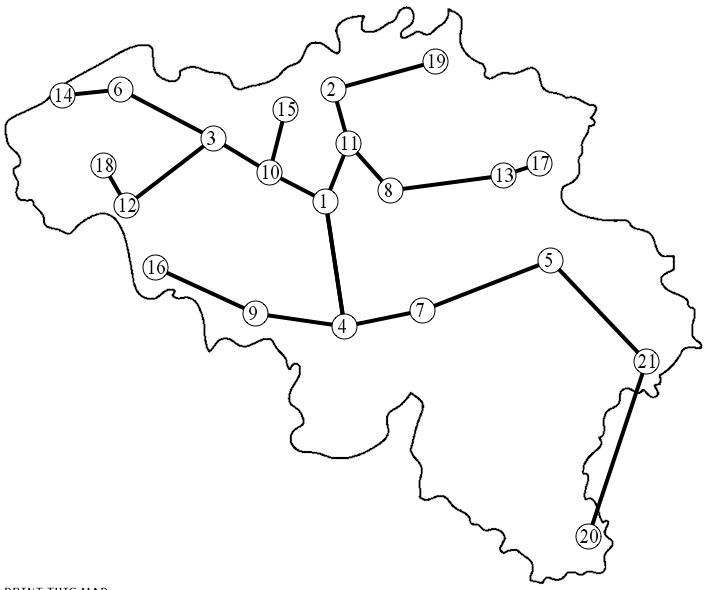}}
\caption{Proximity graphs constructed on sites of ${\mathbf U}$.  (a)~relative neighbourhood graph, (b)~Gabriel graph, (c)~minimum spanning tree rooted in Brussels. }
\label{proximitygraphs}
\end{figure}

 A planar proximity graph is a planar graph where two points are connected by an edge if they are close in some sense. A pair of points is assigned a certain neighbourhood, and points of the pair are connected by an edge if their neighbourhood is empty.  Here we consider the most common proximity graph as follows.
\begin{itemize}
\item $\mathbf{GG}$: Points $a$ and $b$ are connected by an edge in the Gabriel Graph $\mathbf{GG}$ if the 
disc with diameter $dist(a,b)$ centred in middle of the segment $ab$ is empty~\cite{gabriel_sokal_1969, matula_sokal_1984} (Fig.~\ref{proximity}a).
\item $\mathbf{RNG}$: Points $a$ and $b$ are connected by an edge in the Relative Neighbourhood Graph $\mathbf{RNG}$ if no other point $c$ is closer
to $a$ and $b$ than $dist(a,b)$~\cite{toussaint_1980} (Fig.~\ref{proximity}b).
\item $\mathbf{MST}$: The Euclidean minimum spanning tree (MST)~\cite{nesetril} is a connected acyclic graph which has the minimum possible sum of edges' lengths (Fig.~\ref{proximity}b).
\end{itemize}
In general, the graphs relate as
$\mathbf{MST} \subseteq \mathbf{RNG}  \subseteq\mathbf{GG}$~\cite{jaromczyk_toussaint_1992,matula_sokal_1984,toussaint_1980}; 
this is called  the Toussaint hierarchy.

\begin{figure}[!tbp]
\centering
\subfigure[${\mathbf{RNG \cap H}}$]{\includegraphics[width=0.49\textwidth]{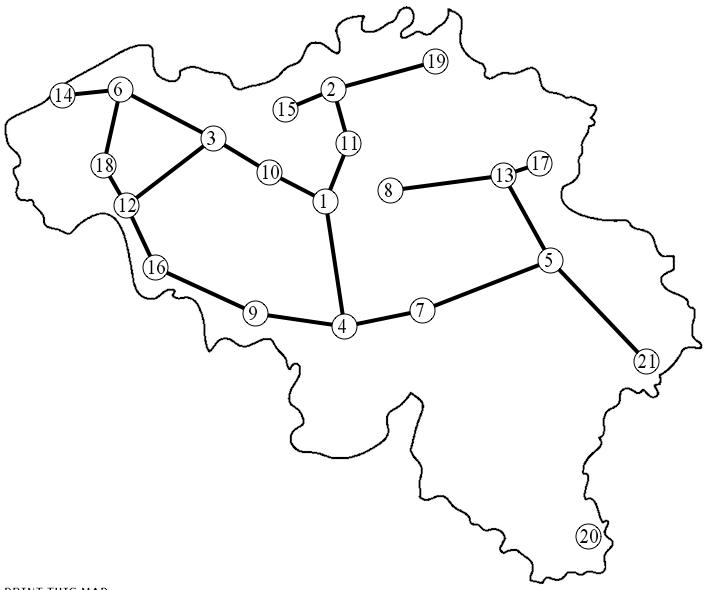}}
\subfigure[${\mathbf{GG \cap H}}$]{\includegraphics[width=0.49\textwidth]{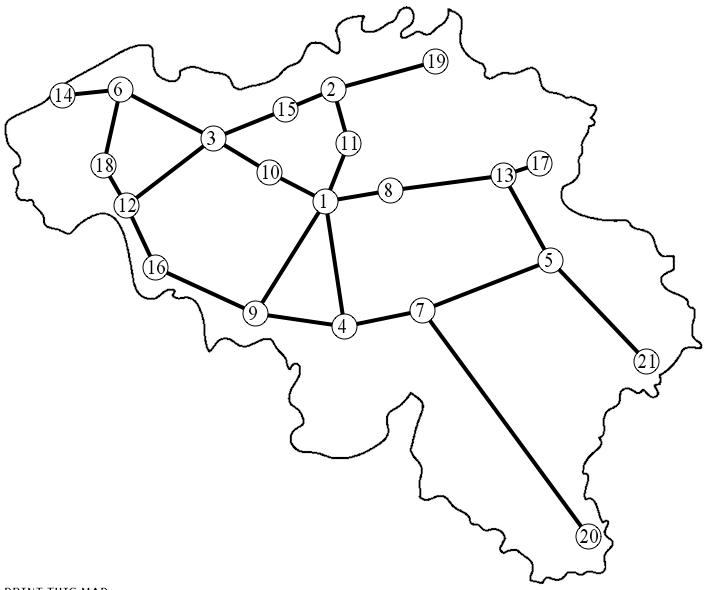}}
\subfigure[${\mathbf{MST(1) \cap H}}$]{\includegraphics[width=0.49\textwidth]{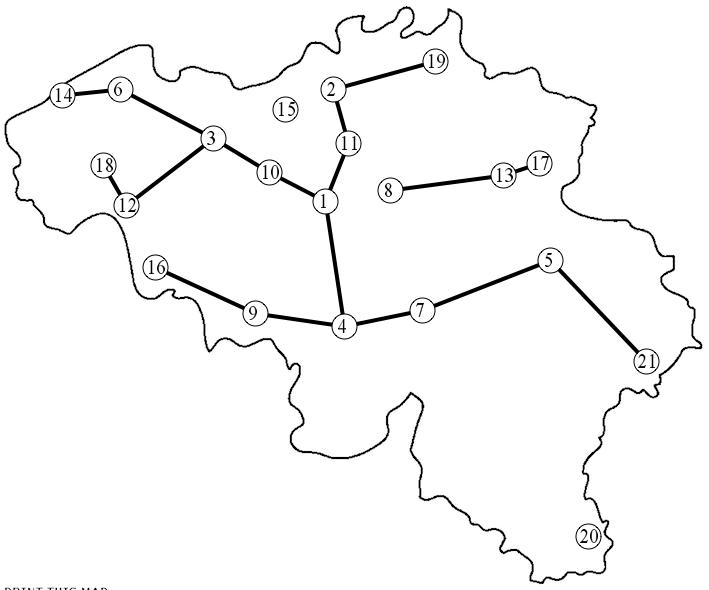}}
\caption{Intersections of the Belgian motorway graph with (a)~relative neighbourhood graph, (b)~Gabriel graph, 
(c)~minimum spanning tree rooted in Brussels.}
\label{motorwaysintersection}
\end{figure}

\begin{finding}
The Belgian motorway graph is approximated by the Gabriel graph with 80\% accuracy and by the relative neighbourhood graph with 
70\% accuracy. 
\end{finding}

The following motorways links from $\mathbf{H}$ are not presented in the Gabriel graph $\mathbf{GG}$: 
(Brussels area --- Antwerp area), ( Brussels area --- Tournai), (Brussels area --- Namur), (Antwerp area --- Hasselt), 
(Leuven --- Li\`{e}ge area), (Li\`{e}ge area --- Arlon) (Fig.~\ref{motorwaysintersection}a).  With regards to the relative neighbourhood graph, 
a few more transport links from $\mathbf{H}$ are not a part of $\mathbf{RNG}$: (Brussels area --- Mons), (Brussels area --- Leuven) and (Namur --- Arlon) (Fig.~\ref{motorwaysintersection}b). Thus $\mathbf{GG}$ represents 23 of 29 edges of the motorway graph $\mathbf{H}$ and $\mathbf{RNG}$ 20 of 29 edges of  $\mathbf{H}$.

\begin{finding}
The minimum spanning tree rooted in Brussels is not a subgraph of the Belgian motorways graph, $\mathbf{MST} \nsubseteq \mathbf{H}$. 
\end{finding}

This is because the links (Mechelen --- Sint-Niklaas) and (Mechelen --- Leuven) exist in  $\mathbf{MST}$ but not are not present in $\mathbf{H}$ (Fig.~\ref{motorwaysintersection}c). The minimum spanning tree is considered to be an optimal, 
in a sense of minimality of edge lengths, acyclic planar graph. The fact that two of the spanning tree edges are not represented by man-made motorway links allows us to suggest that the Belgian motorway network is not optimal, at least not optimal in spanning of major urban areas.

\begin{figure}[!tbp]
\centering
\subfigure[${\mathbf{RNG \cap P}(\frac{11}{28})}$]{\includegraphics[width=0.49\textwidth]{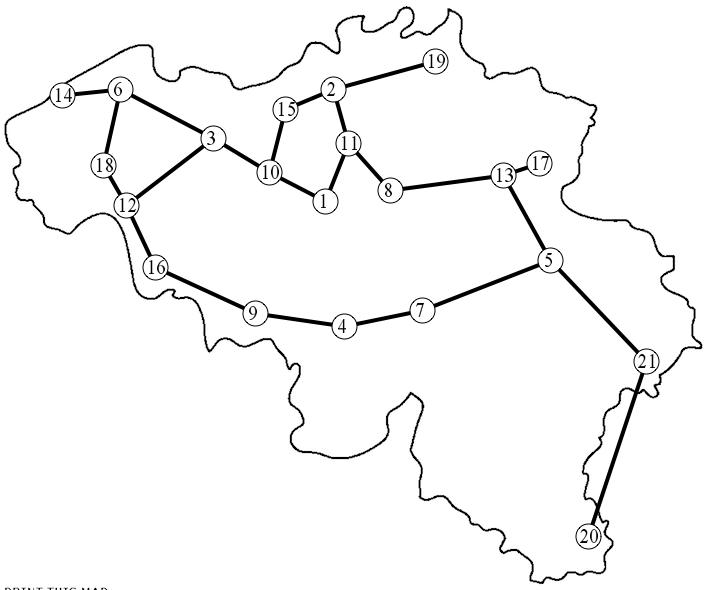}}
\subfigure[${\mathbf{GG \cap P}}(\frac{11}{28})$]{\includegraphics[width=0.49\textwidth]{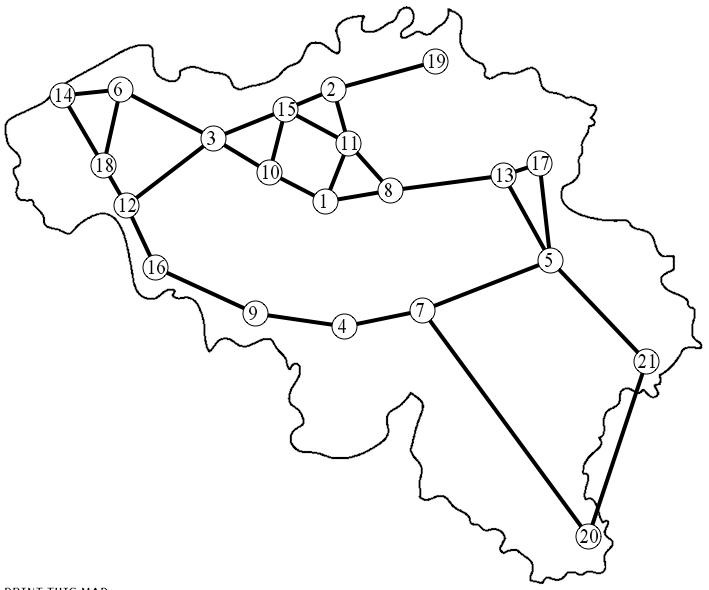}}
\subfigure[${\mathbf{MST \cap P}}(\frac{11}{28})$]{\includegraphics[width=0.49\textwidth]{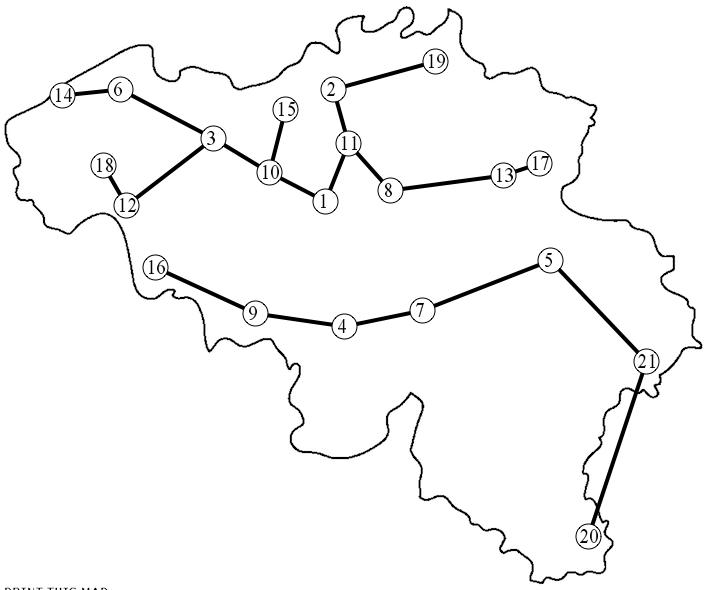}}
\caption{Intersections of the Physarum graph $\mathbf{P}(\frac{11}{28})$ with 
(a)~relative neighbourhood graph, (b)~Gabriel graph, 
(c)~minimum spanning tree rooted in Brussels.}
\label{physarum11_intersectproximitygraphs}
\end{figure}

Considering Physarum graph $\mathbf{P}(\frac{11}{28})$ is credible because its edges are represented by protoplasmic tubes in at least 40\% of laboratory experiments, we can compare it with three basic proximity 
graphs (Fig.~\ref{physarum11_intersectproximitygraphs}).

\begin{finding}
If slime mould represented A7/A54 Brussels --- Charleroi, non-existing in reality motorways between 
Leuven --- Namur and Turnhout --- Hasselt its credible transport network would be a super-graph of the relative neighbourhood graph. 
\end{finding}

This is a direct outcome of  comparing Fig.~\ref{physarum11_intersectproximitygraphs}a and 
Fig.~\ref{proximitygraphs}a. Relative neighbourhood graphs are considered to be optimal cyclic graphs in terms of total edge length and travel distance, and are known to be a good approximation of road networks~\cite{watanabe_2005,watanabe_2008}. The Leuven --- Namur and Brussels --- Charleroi links are represented by slime mould in less than 20\% of laboratory experiments (Fig.~\ref{selectedphysarumgraphs}b). The intersection of the Physarum graph  $\mathbf{P}(\frac{11}{28})$ (Fig.~\ref{selectedphysarumgraphs}d) with the minimum spanning tree (Fig.~\ref{proximitygraphs}c) comprises two disconnected components: one lies in Flanders and another in Wallonia (Fig.~\ref{physarum11_intersectproximitygraphs}c).

\section{Dissolution: Transport networks and administrative subdivision}
\label{adminstrativesubdivision}

The province Brabant Wallone has no transport routes originating in and passing through it for $\theta=\frac{10}{28}$.
When $\theta$ increases to $\frac{11}{28}$, the Antwerp province becomes isolated from the other provinces of Belgium. For 
$\theta=\frac{22}{28}$ we have isolated Antwerp and Limburg provinces, and clusters of interconnected regions: 
\begin{enumerate}
\item West-Vlaanderen, Hainaut and Namur provinces,
\item West-Vlaanderen, Oost-Vlaanderen, Vlaams Brabant, Oost-Vlaanderen provinces,
\item Li\`{e}ge and Luxembourg provinces.
\end{enumerate}

\begin{figure}[!tbp]
\centering
\includegraphics[width=0.49\textwidth]{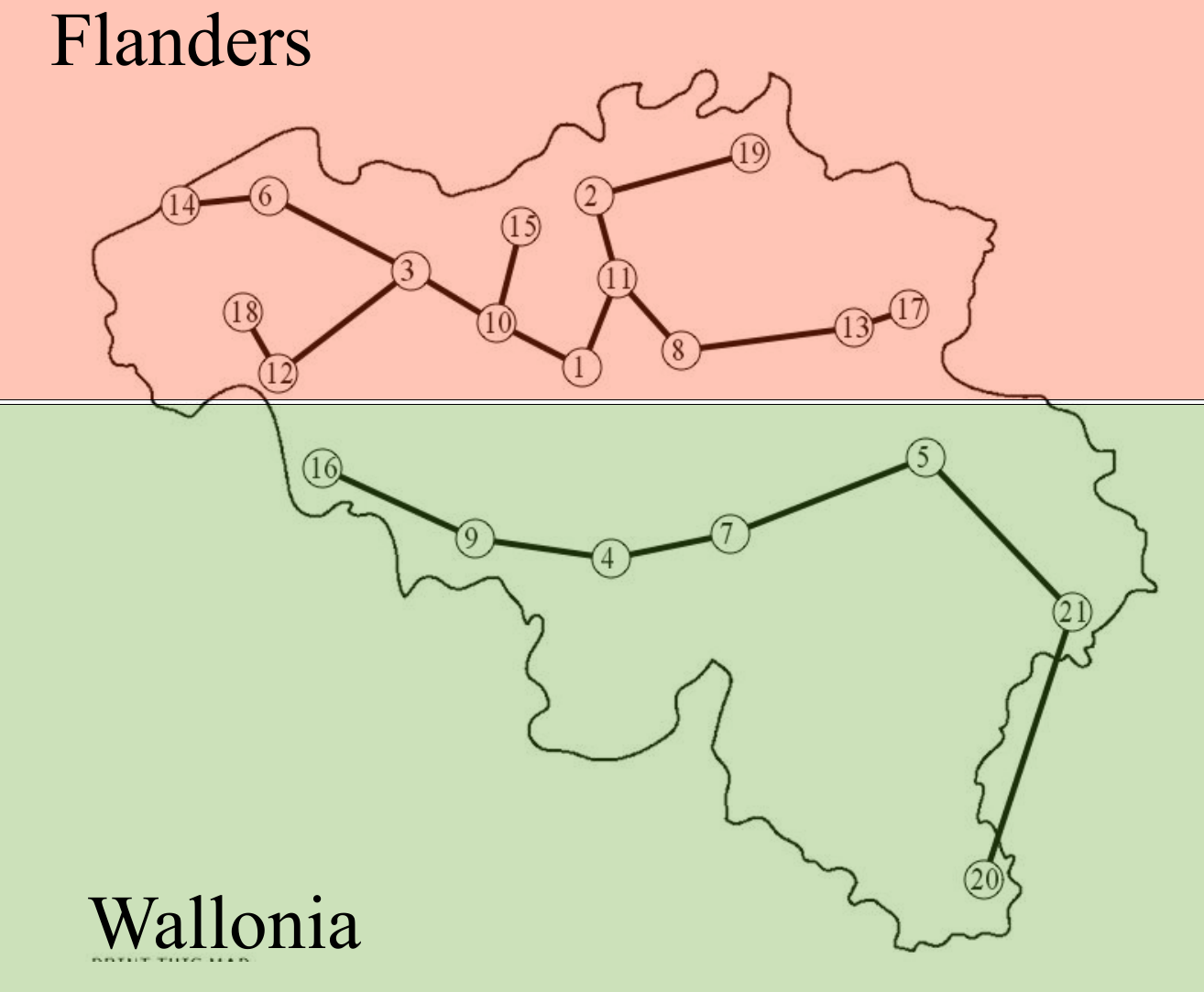}
\caption{Intersection of Physarum graph $\mathbf{P}(\frac{11}{28})$ with the minimum spanning tree rooted in Brussels.}
\label{walloniaflanders}
\end{figure}

In terms of spanning tree (Fig.~\ref{proximitygraphs}) the transport link between the Brussels area and the Charleroi area is the only means of keeping the Dutch and French speaking communities together. 

\begin{finding}
If the two parts of Belgium were separated with Brussels in Flanders, the Walloon region of the Belgian transport network would be represented by a single chain from Tournai in the north-west to the Li\`{e}ge area in the north-east and down to southmost Arlon. 
\end{finding}

This transport link is not part of the  spanning tree rooted in Brussels, and when omitted the two Belgian communities become isolated from each other (Fig.~\ref{walloniaflanders}). 

\section{Response of slime mould networks to imitated disasters}
\label{contamination}

\begin{figure}[!tbp]
\centering
\subfigure[]{\includegraphics[width=0.3\textwidth]{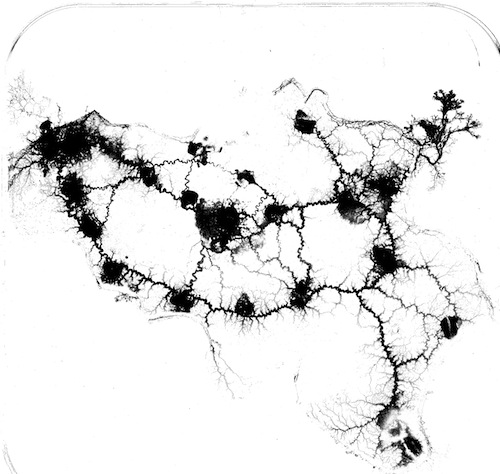}}
\subfigure[]{\includegraphics[width=0.3\textwidth]{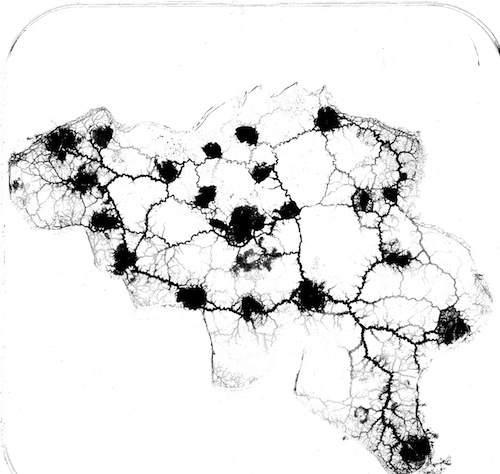}}
\subfigure[]{\includegraphics[width=0.3\textwidth]{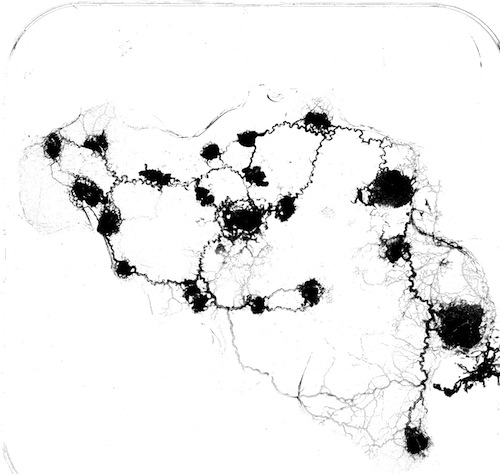}}
\subfigure[]{\includegraphics[width=0.3\textwidth]{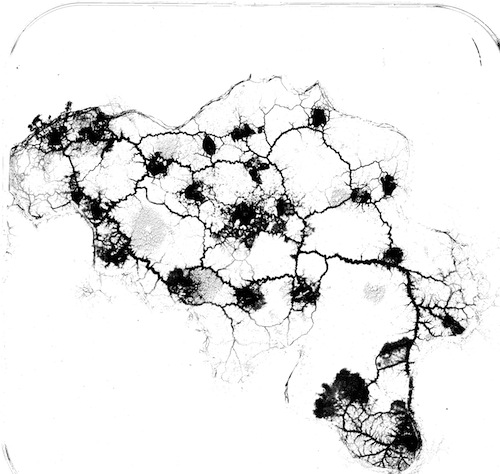}}
\subfigure[]{\includegraphics[width=0.3\textwidth]{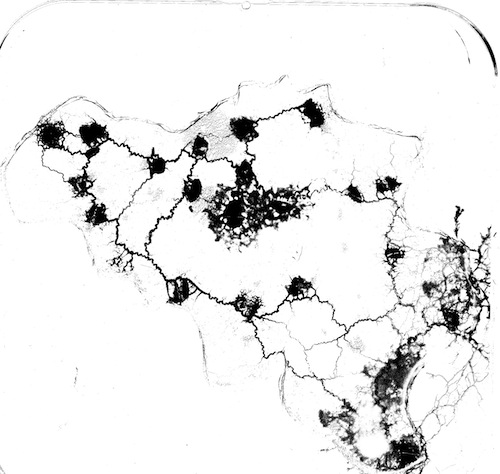}}
\subfigure[]{\includegraphics[width=0.3\textwidth]{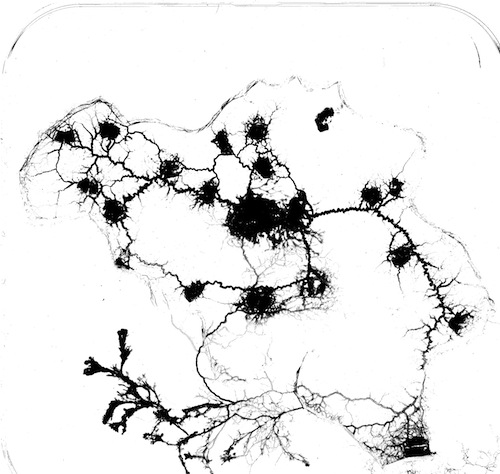}}
\subfigure[]{\includegraphics[width=0.3\textwidth]{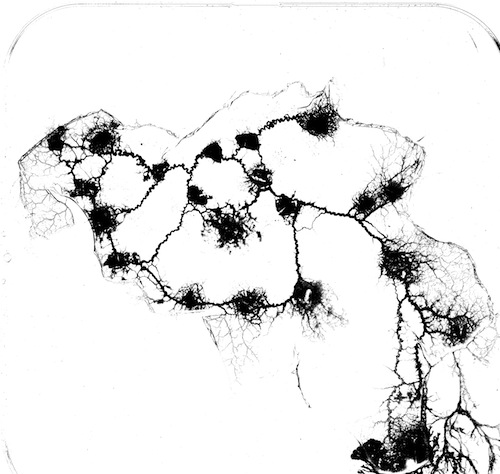}}
\caption{Response to spreading contamination recorded in laboratory experiments. Epicentre of contamination is in  Antwerp area.}
\label{saltat2}
\end{figure}

\begin{figure}[!tbp]
\centering
\subfigure[]{\includegraphics[width=0.3\textwidth]{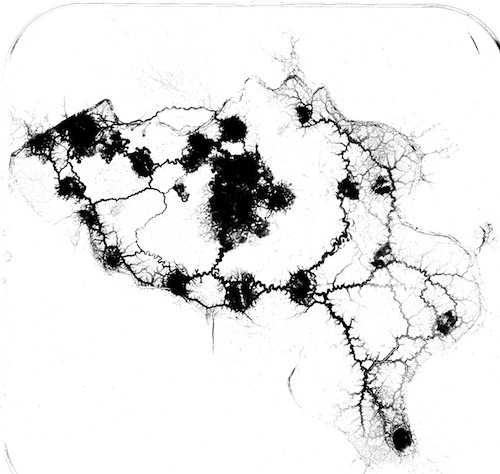}}
\subfigure[]{\includegraphics[width=0.3\textwidth]{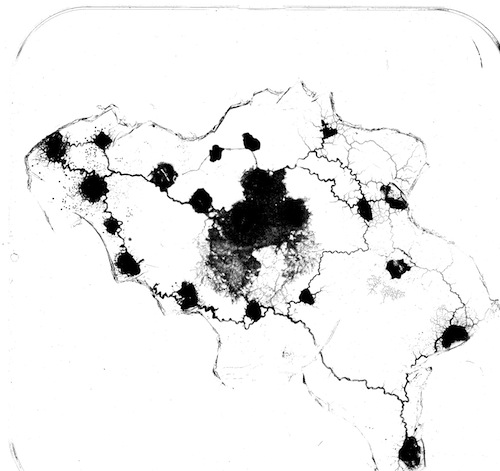}}
\subfigure[]{\includegraphics[width=0.3\textwidth]{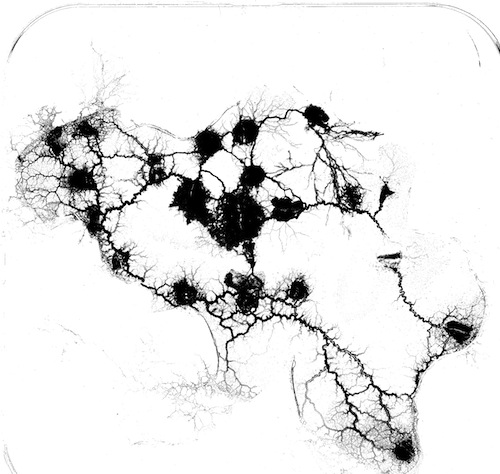}}
\subfigure[]{\includegraphics[width=0.3\textwidth]{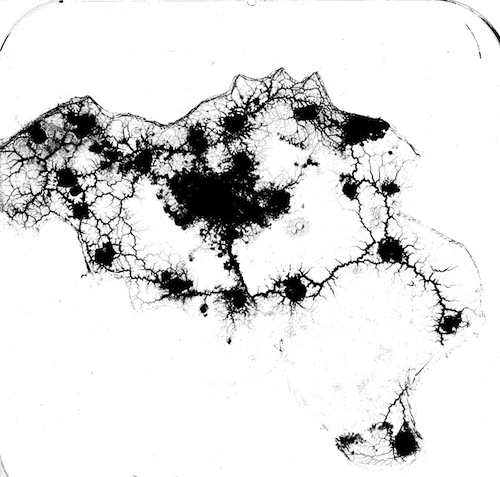}}
\subfigure[]{\includegraphics[width=0.3\textwidth]{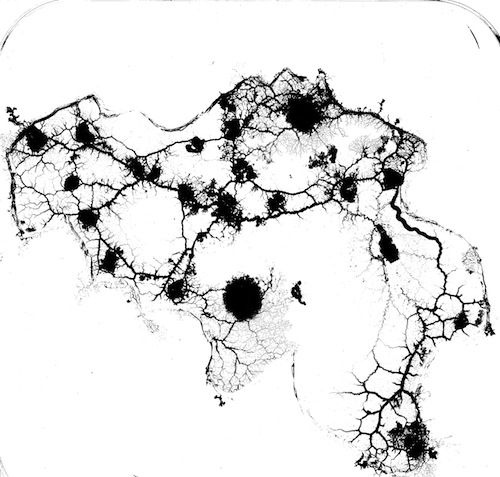}}
\subfigure[]{\includegraphics[width=0.3\textwidth]{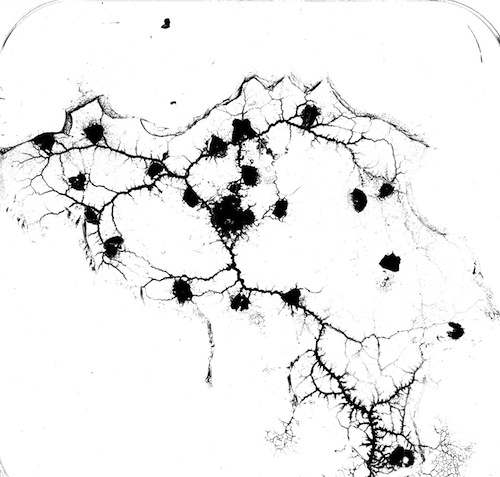}}
\subfigure[]{\includegraphics[width=0.3\textwidth]{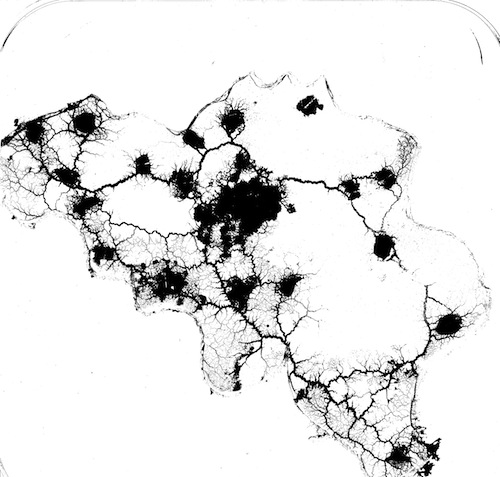}}
\subfigure[]{\includegraphics[width=0.3\textwidth]{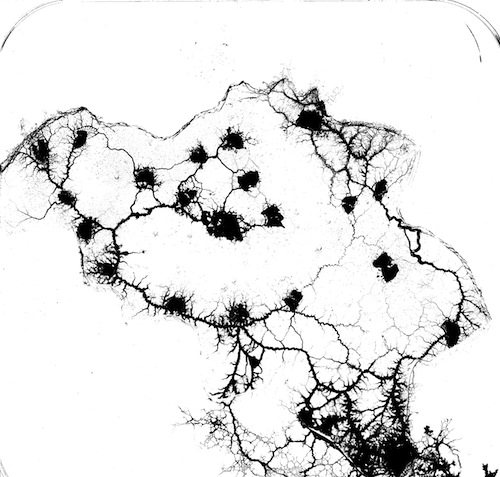}}
\subfigure[]{\includegraphics[width=0.3\textwidth]{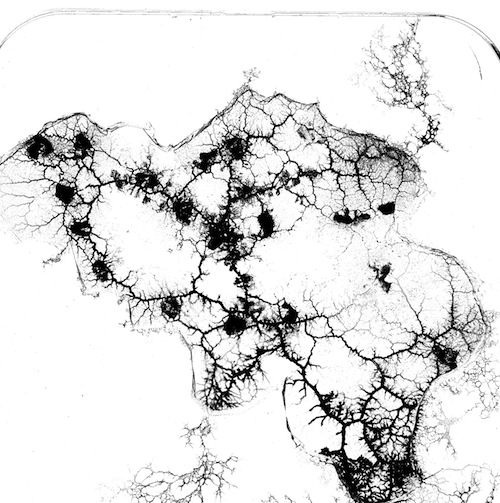}}
\caption{Response to spreading contamination recorded in laboratory experiments. Epicentre of contamination is in 
Li\`{e}ge area.}
\label{saltat5}
\end{figure}

To study the  reaction of Physarum-grown transport networks on major disasters, we placed crystals of sodium chloride in the approximate positions of Doel Nuclear Power Plant, near Antwerpen (seven experiments) and Tihange Nuclear Power Station, near Li\`{e}ge (nine experiments).  Sodium chloride is a strong repellent for \emph{P. polycephalum}. The salt diffuses in the  substrate outwards its original application site (epicentre of disaster). It can therefore imitate radioactive and/or chemical pollution and subsequent disturbance spreading along Belgian transport networks. Images of protoplasmic networks reconfigured 24~h after start of contamination are shown in Figs.~\ref{saltat2} and \ref{saltat5}.

\begin{figure}[!tbp]
\centering
\includegraphics[width=0.95\textwidth]{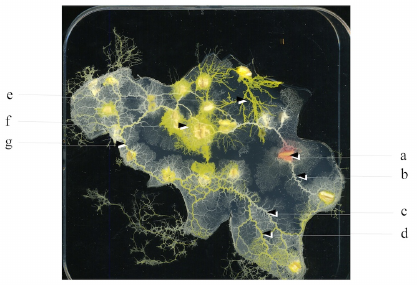}
\caption{Illustration of the slime mould's response to contamination. 
(a)~epicentre of contamination. 
(b)~abandoned and not repositioned transport link. 
(c)~abandoned and repositioned transport link. 
(d)~new location of the abandoned transport link. 
(e)~explorative activity.
(f)~emergency preparations, initial stage of sclerotinisation.
(g)~enhanced transport links. Image is taken 24~h after initiation of contamination in Li\`{e}ge area.}
\label{saltexample}
\end{figure}

A typical response to propagating contamination is dissected in Fig.~\ref{saltexample}. In this example we place a salt grain on the oat flake representing 
the Li\`{e}ge area  (Fig.~\ref{saltexample}, 'a'). In 24~h after the imitated accident, the contamination spreads as far Namur on the west and Sankt-Vith on the east and Hasselt and Genk in the south. Transport links in proximity of the contamination epicentre become destroyed and abandoned. An example of abandoned transport link is a protoplasmic tube (marked 'b' Fig.~\ref{saltexample}) representing E42/A27 (Li\`{e}ge to Sankt-Vikt motorway).  No alternative routes for such destroyed transport links are offered by the slime mould.  

Transport links being at a significant distance from the epicentre but yet inside the contamination zone are shifted further away from the contamination. For example, slime mould abandons protoplasmic tube representing motorway E411, marked 'c' in Fig.~\ref{saltexample} but grows another tube, marked 'd' in Fig.~\ref{saltexample} slightly westwardly. 
Urban areas not directly affected by contamination show signs of increased explorative and scouting activity, for example,
Antwerp and Turnhout areas, marked 'e' in Fig.~\ref{saltexample}. Also transport links not affected by contamination became visibly enhanced (Fig.~\ref{saltexample}, 'g'). In the situation of continuing contamination plasmodium 
'considers' the last opportunity to survive by forming sclerotium (hardened body of the 'hibernating' slime mould), initial stage of the sclerotium formation is labelled 'f' in Fig.~\ref{saltexample}.

Based on outcomes of our scoping experiments (Figs.~\ref{saltat2} and \ref{saltat5}) we can propose the following scenarios of response to contaminations:
\begin{itemize}
\item Epicentre of contamination is at the Tihange Nuclear Power Station near Li\`{e}ge. Segments of transport network connecting Namur,  Li\`{e}ge, Hasselt, Genk, Sankt-Vith and Arlon are destroyed or abandoned. Hyperactivity is observed in domains surrounding Sinkt-Niklaas, Antwerp and Turnhout. Preparations for emergency hibernation take place at 
Aalst, Brussels, Leuven and Mechelen.  Transport links between Gent, Roeselare, Kortijk, Tournai, Mons, Charleroi, Namur
are hypertrophied (Fig.~\ref{saltat5}).
\item Epicentre of contamination is at the Doel Nuclear Power Plant near Antwerpen. Domains surrounding Antwerp, Sinklaas, Mechelen, Turnhout become depopulated and transport links are abandoned. Explorative activity is observed in areas of Oostende, Brugge, Roeselare, and Arlon and Sankt-Vith. Attempted mass-migration is recorded into Northern France, South of the Netherlands and Luxemburg and West Germany. Transport links in the chain Brugge --- Roeselare --- Kortijk --- Tournai --- Mons --- Charleroi --- Namur --- Li\`{e}ge --- Sankt-Vith/Arlon are significantly  hypertrophied
(Fig.~\ref{saltat2}).  Preparations for emergency hibernation take place at Brussels and Leuven, and at Hasselt and Genk areas. 
\end{itemize}

\section{Conclusions}

To evaluate how good  Belgian motorways are from an amorphous living creature point of view, we conducted the following laboratory experiments with slime mould \emph{P. polycephalum}. We represented major urban areas with oat flakes and inoculated slime mould in the oat flake corresponding to Brussels. We waited till the slime mould colonised all oat flakes and then analysed the slime mould's protoplasmic network structure and compared it with 
the man-made motorway network and basic planar proximity graphs.  We found that \emph{P. polycephalum} 
almost but not completely approximates  the Belgian motorway network. Transport links 
Roeselare --- Kortrijk --- Mons --- Charleroi --- Namur,  
Oostende --- Brugge ---  Gent ---  Aalst --- Brussels --- Leuven --- Mechelen --- Antwerp --- Sint-Niklaas,
Li\`{e}ge --- Sankt-Vith --- Arlon, and Hasselt ---  Genk appear as protoplasmic tubes in almost all laboratory experiments.  
Motorway links connecting Brussels with Antwerp, Tournai, Mons, Charleroi, Namur, and links connecting Leuven with Li\`{e}ge and Antwerp with Genk and Turnhout are ``considered'' by slime mould to be redundant  and thus almost never appear in our experiments with the slime mould. If slime mould represented A7/A54 Brussels --- Charleroi, non-existing in reality, motorways between Leuven --- Namur and Turnhout --- Hasselt its credible transport network would be a super-graph of the relative neighbourhood graph.

Motorway segments A17 (Antwerp --- Sint-Niklaas), A19 (Antwerp --- Mechelen),  E42 (Li\`{e}ge --- Sankt-Vith), 
A10/E40 (Oostende --- Gent --- Aalst --- Brussels --- Leuven), A17/E403 (Roeselare --- Kortrijk --- Tournai), E42/E19 (Tournai --- Mons), E42 (Mons --- Charleroi --- Namur) are essential transport links from the slime mould's point of view. 
If the two parts of Belgium were separated with Brussels in Flanders, the Walloon region of the Belgian transport network would be represented by a single chain from Tournai in the north-west to Li\`{e}ge area in the north-east and down to southmost Arlon.  

While imitating major disasters leading to contamination propagating from Li\`{e}ge and Antwerp areas we found several scenarios of major restructuring of transport networks and possible routes of mass-migration. 

We believe the results of our scoping experiments and analysis can be used not only in bio-inspired unconventional 
computing but also in more classical fields of urban and transport planning. Possible applications include but are not limited
to restructuring of rural landscape~\cite{froment:1987}, novel approaches towards mapping vehicular accessibility~\cite{Vandenbulcke:2009},  simulation of traffic dynamics in Belgian motorways~\cite{Boel:2006}, 
development of Trans-European transport networks~\cite{De Lathauwer:1995}, bottom up landscape planning~\cite{Sevenant:2010} and modelling relations between urban sprawl and growing transport networks~\cite{Poelmans:2009}.

\newpage

\section{Appendix A: Physarum graphs for all values of $\theta$}

\end{document}